\documentclass[manuscript, nonacm]{acmart}

\AtBeginDocument{%
  }

\usepackage{booktabs}
\usepackage{tabularx}
\usepackage{pifont} 

\definecolor{deeppink}{rgb}{1.0, 0.08, 0.58}
\definecolor{cadmiumgreen}{rgb}{0.0, 0.42, 0.24}
\usepackage[table]{xcolor}

\usepackage{subcaption}
\usepackage{tikz}
\usepackage{pgfplots}
\usepackage{pgf-pie}
\pgfplotsset{compat=1.18}
\usepgfplotslibrary{groupplots}

\usepackage{multirow}
\usepackage{todonotes}

\newcommand{\ourtool}{PersuLab}

\usepackage{url}
\begin{document}

\title{Games That Teach, Chats That Convince: Comparing Interactive and Static Formats for Persuasive Learning}

\author{Seyed Hossein Alavi}
\email{salavis@cs.ubc.ca}
\affiliation{
\institution{University of British Columbia}
\country{Canada}
}
\affiliation{%
  \institution{Vector Institute for AI}
  \country{Canada}
}

\author{Zining Wang}
\affiliation{
\institution{University of British Columbia}
\country{Canada}
}
\affiliation{%
  \institution{Vector Institute for AI}
  \country{Canada}
}

\author{Shruthi Chockkalingam}
\affiliation{
\institution{University of British Columbia}
\country{Canada}
}

\author{Raymond T. Ng}
\affiliation{
\institution{University of British Columbia}
\country{Canada}
}

\author{Vered Shwartz}
\affiliation{
\institution{University of British Columbia}
\country{Canada}
}
\affiliation{%
  \institution{Vector Institute for AI}
  \country{Canada}
}

\begin{abstract}
Interactive systems such as chatbots and games are increasingly used to persuade and educate on sustainability-related topics, yet it remains unclear how different delivery formats shape learning and persuasive outcomes when content is held constant. Grounding on identical arguments and factual content across conditions, we present a controlled user study comparing three modes of information delivery: static essays, conversational chatbots, and narrative text-based games. Across subjective measures, the chatbot condition consistently outperformed the other modes and increased perceived importance of the topic. However, perceived learning did not reliably align with objective outcomes: participants in the text-based game condition reported learning less than those reading essays, yet achieved higher scores on a delayed (24-hour) knowledge quiz. Additional exploratory analyses further suggest that common engagement proxies, such as verbosity and interaction length, are more closely related to subjective experience than to actual learning. These findings highlight a dissociation between how persuasive experiences feel and what participants retain, and point to important design trade-offs between interactivity, realism, and learning in persuasive systems and serious games.
\end{abstract}


\received{18 February 2026}

\maketitle

\section{Introduction}
\label{sec:introduction}

Persuasive and educational technologies have been widely explored for environmental sustainability, often aiming to raise awareness, shift attitudes, or encourage pro-environmental behavior \citep{brynjarsdottir2012sustainably,Adaji2022Review}. 
Recent advances in large language models (LLMs) have further accelerated this trend by reducing the barriers for developing conversational agents and interactive games. 
Enabling dynamically generation of persuasive content at scale makes these interactive systems alternatives preferable to traditional static formats such as essays or informational webpages.

Despite this growing adoption, there remains limited empirical understanding of how different \emph{modes of delivery} shape persuasive experience and learning outcomes when the underlying informational content is held constant. Many evaluations of persuasive and educational systems rely heavily on self-reported outcomes such as engagement, enjoyment, perceived learning, or attitude change \citep{yang2023does, de2017persuasive}. While these measures capture important aspects of user experience, prior work has shown that subjective perceptions of learning do not always align with objectively measured learning or longer-term knowledge retention \citep{Persky2020PerceptionOL,AlAperaparSingh2023THEDO}. This misalignment is particularly relevant for interactive systems, where engagement and interactivity may shape how effective an experience \emph{feels} without necessarily improving what users ultimately retain.

This question is especially relevant for designers of games-for-change and serious games, which are widely used for sustainability education and persuasion \citep{Monroe03062019,Newsome2020,Fernandez_Galeote_2021}. While interactive narratives can increase engagement, they may also introduce trade-offs in realism, trust, or cognitive load that affect persuasive impact \citep{Apple2007,Appel2013,doi:10.1177/0093650212450585}. In contrast, static formats such as essays may appear clearer or more credible but encourage passive consumption. \emph{Controlled comparisons} are therefore needed to disentangle the effects of \emph{content} from \emph{interaction structure}.

In this work, we address this gap through a controlled comparison of three common information delivery modes: a static essay, a conversational chatbot, and a narrative-driven text-based game. Crucially, all three conditions were grounded in the \emph{same} set of persuasive arguments and numerical facts, allowing us to examine how delivery format alone shapes subjective experience, perceived attitude change, and objective knowledge retention. We focus on two sustainability-related topics -- recycling and public transit -- chosen for their real-world relevance and suitability for persuasion-oriented interventions.

We conducted a between-subjects user study in which participants experienced exactly one delivery mode and one topic. We collected pre-study measures, post-study subjective ratings and perceived change measures, and administered a delayed (24-hour) objective knowledge quiz. We also analyze the interaction logs for the interactive conditions. This combination enables us to examine not only differences across modes, but also the relationship between interaction behavior, subjective impressions, and learning outcomes.

Our findings reveal several important tensions. Conversational interaction consistently improved subjective experience and increased perceived importance of the topic, but perceived learning does not reliably predict objective retention: participants in the text-based game condition reported learning less, yet retained more factual information than those who read an essay. Exploratory analyses of interaction logs further suggest that common engagement proxies -- such as verbosity, turn count, or session duration -- are more strongly associated with subjective impressions than with actual learning. Together, these results highlight a dissociation between how persuasive experiences \emph{feel} and what participants ultimately \emph{retain}. 
  
By clarifying the trade-offs between interactivity, realism, subjective experience, and learning, this work aims to inform the design and evaluation of future persuasive chatbots, narrative games, and games-for-change.

\section{Background}
\label{sec:background}

Given the abundance of research on interactive technologies for educational and persuasive purposes, we primarily focus our discussion on technologies developed for the topic of sustainability and climate change. We describe various types of systems and what makes them effective (\S\ref{sec:bg:interactive}), discuss ways to measure learning in such systems (\S\ref{sec:bg:retention}), and finally more broadly discuss persuasive technologies (\S\ref{sec:bg:persuasive}). 

\subsection{Educational Technology for Sustainability-Related Topics}
\label{sec:bg:interactive}

Environmental sustainability is a prominent topic in HCI research and is frequently addressed through persuasive approaches~\cite{brynjarsdottir2012sustainably}. 
Persuasive techniques have been applied in educational technologies, where they are used to help learners form, alter, or reinforce attitudes and behaviors that support learning~\cite{app112110089}. Under this framing, two interactive modalities have emerged as particularly effective for promoting sustainability-related learning and persuasion.

Conversation is especially effective at persuading people \cite{mercier2019enigma,mercier2020notborn}. Accordingly, conversational agents (or chatbots) have emerged as a simple, widely used interactive modality, which is especially popular in recent years with the advent of LLMs. Prior work has examined the role of chatbots in persuasion \cite{Kobi2025} as well as in educational contexts \cite{hobert2019say,okonkwo2020python}. LLM-based chatbot engaging in conversation might be substantially more persuasive than single-message interactions. They can promote more active engagement, quickly address user concerns, and tailor arguments to individual preferences \cite{pnasnexus,matz2024potential}. When designed responsibly, chatbots can serve as enablers of positive individual and social change \cite{furumai-etal-2024-zero,chockkalingam-etal-2025-go}. 

An alternative interactive technology is serious games: games that are designed for educational rather than entertainment purposes \cite{nazry2017mood}. The interactivity of serious games makes them one of the most effective strategies for providing educational content in an engaging manner \cite{Monroe03062019}. Serious games are thus a popular approach to teach people about sustainability \cite{Newsome2020,fernandez2021,chisalita2022,Troiano2020}. Numerous empirical studies demonstrate that games and gamification are effective in engaging with climate change in educational contexts \cite{Fernandez_Galeote_2021}.  For example, digital games has been demonstrated to improve students' climate literacy as well as their behaviors towards energy related topics \cite{HARKERSCHUCH2020103705,Nussbaum2015, Newsome2020,knol2011enercities}. 

In particular, narrative is a common component of climate change games \cite{fernandez2021}. Serious games that presented a story in a virtual location based on a real city increased players' concern about climate issues\cite{Angel2015}. This is also our focus in this paper; for the game interactive modality, we specifically look at a text-based narrative games.

Despite this body of work, relatively little research has systematically compared how different interactive systems -- such as games and conversational agents -- differ from one another in both learning and persuasion outcomes, which we do in this work. 

\subsection{Measuring Learning in Interactive Systems}
\label{sec:bg:retention}

Prior work on interactive learning system uses both subjective self-reports of perceived learning and objective performance-based assessments, reflecting different assumptions about what constitutes learning and how it should be evaluated. 
Some studies operationalize learning outcomes as perceived learning effects, capturing the extent to which learners experience an interaction as educational \cite{yang2023does}. In contrast, others adopt objective evaluation strategies, assessing learning through changes in knowledge or performance \cite{Nussbaum2015}. The choice of evaluation metrics can lead to divergent conclusions about system effectiveness. 

For example, \citeauthor{Nussbaum2015} \cite{Nussbaum2015} developed a serious game exploring water-level decline and evaluated learning using pre-study, post-study, and 11-day delayed post-study tests to measure the immediate and long-term knowledge retention. Participants who played the game demonstrated effective long-term knowledge retention, as evidenced by maintained marginal gains on both the post-study and delayed post-study test compared to a control group that engaged in reading a static website. 

While our work reaffirms this conclusion, it also shows that games underperform in perceived learning metrics. Our work aligns with prior research showing that subjective perceptions of learning may not align with objectively measured learning outcomes in interactive educational systems. For instance, \citeauthor{Persky2020PerceptionOL} \cite{Persky2020PerceptionOL} found no correlation between perceived and actual knowledge gains among 277 college and pharmacy students. Similarly, a study comparing perceived and actual academic performance in an online learning environment with 382 participants found no significant relationship between the two measures \cite{AlAperaparSingh2023THEDO}. This misalignment highlights the need to evaluate interactive learning systems using both subjective and objective measures to better capture true learning outcomes.

\subsection{Persuasive Technology for Sustainability}
\label{sec:bg:persuasive}

Given that our study focuses on sustainability issues, it ties together both learning about the topic and belief change. While we primarily focus on learning outcomes, we also review relevant literature on persuasive technology. \citeauthor{de2017persuasive} \cite{de2017persuasive} defines persuasive technology as ``technologies aimed at changing people's attitudes or behaviors through persuasion and social influence, but not through coercion or deception''. 

Classic persuasive technology interventions for sustainability typically aim to raise awareness, personalize interventions, and target specific behavior changes \cite{Adaji2022Review}. Such technologies have been implemented across a wide range of media, including desktop applications \cite{Adaji2020EvaluatingTS}, mobile applications~\cite{matthews2016persuasive}, and serious games \cite{su141610400,folkins2020}. Among these, serious games have gained particular popularity within sustainability-focused persuasive technology research, promoting environmental sustainability goals such as energy conservation and waste reuse \cite{Adaji2022Review}. 

Recent advances in LLMs have shifted some of the focus towards chatbots. LLMs can engage in sophisticated interactive dialogue, making them powerful tools for shaping attitudes, preferences, and behaviors \cite{Kobi2025,durmus2024persuasion,salvi2025conversational,matz2024potential}. In particular, their ability to deliver adaptive and personalized persuasive content at unprecedented scale makes LLM-based systems \emph{hyper-persuasive technologies} \cite{luciano2024hypersuasion}. LLMs' persuasive power has been demonstrated across multiple domains, including consumer marketing \cite{nam2025llms}, healthcare \cite{karinshak2023working,altay2023information}, politics \cite{goldstein2024persuasive, bai2025llm,fernbach2013political,pnasnexus}, life-style decisions \cite{chockkalingam-etal-2025-go}, and pro-environmental appeals \cite{matz2024potential}.

Multiple studies found LLM-generated content was perceived as at least as persuasive, and often more persuasive than human-written content \cite{karinshak2023working,matz2024potential,bai2025llm,chockkalingam-etal-2025-go}. Other studies have investigated factors that influence LLM persuasiveness, including prompt design and message characteristics \cite{rogiers2024persuasion,pnasnexus,altay2023information,Kobi2025}. In particular, evidence-based persuasion -- where models structure dialogue around fact-checkable, argument-relevant information -- has been shown to outperform approaches that rely primarily on rhetorical or emotional language \cite{pnasnexus,Kobi2025}. LLM-based serious games are capable of all of the above and are also highly conducive to personalization such as personalized NPCs \cite{alavi2024mcpdial,elmgadmi2025large} and narratives \cite{alavi2024game,shields2025could}, making them effective in persuasion. 
 
LLMs' persuasiveness is typically evaluated against human persuaders \cite{salvi2024conversational,bai2025llm,schoenegger2025large}. LLM-based systems with particular mechanisms for increasing persuasiveness are often compared against other LLM-based baselines \cite{furumai-etal-2024-zero}. Finally, some studies compare interactive LLM-based systems against less interactive settings such as reading static LLM-generated passages \cite{altay2023information,bai2025llm}. However, there remains limited understanding of how LLM persuasion outcomes differ across interaction modalities. Our work addresses this gap by systematically comparing persuasive effects across three modalities that vary in their level and form of interactivity: essays, chatbots, and serious games. 
\section{Study Design}
\label{sec:study-design}
This study examines how information delivery modality influences knowledge retention and persuasive effectiveness in sustainability education. We focus on two topics pertaining to sustainability (\S\ref{sec:study-design:topics}). We conducted a between-subjects user study to compare the knowledge retention and persuasive effectiveness of three information delivery modes: \textbf{essay}, \textbf{chatbot}, and a \textbf{text-based game} (\S\ref{sec:study-design:exp-conditions}). While the three conditions differed in their interaction style and delivery format, all were implemented within a shared system (\ourtool{}) designed to ensure consistent content exposure and comprehensive interaction logging (\S\ref{sec:system}). Each participant (\S\ref{sec:study:participants}) experienced exactly one mode and one topic. They were asked to complete a pre-study questionnaire before the assigned experience and a post-study questionnaire and delayed knowledge assessment afterwards, as we describe in the next section. 

\subsection{Topics, Arguments, and Facts}
\label{sec:study-design:topics}

Participants were assigned to one of two topics: \textbf{recycling} and \textbf{public transit}.  We selected these topics because they represent common, personally actionable sustainability behaviors that are frequently targeted by educational and persuasive technologies \cite{10.1145/2207676.2208394}. For each topic, we used a fixed set of five arguments in favor of taking actions towards sustainability, each paired with a supporting factual statement (Table~\ref{tab:arguments-facts}). Topic assignment was balanced across experimental conditions. To ensure internal validity and enable fair comparison across delivery modes, we strictly controlled informational exposure: in all conditions, the system ensured that all five predefined facts were presented before the interaction could be completed (see Appendix~\ref{appendix:key-arguments} for more details).  


\begin{table*}[t]
\centering
\scriptsize
\setlength{\tabcolsep}{3pt}
\begin{tabularx}{\textwidth}{p{1.2cm} X p{6cm}}
\toprule
\textbf{Topic} & \textbf{Argument (without the premise)} & \textbf{Supporting fact used in the study} \\
\midrule

\multirow{10}{1.2cm}{\textbf{Recycling}} &
(1) Environmental Preservation: Protecting Natural Resources and Ecosystems &
 Recycling one ton of paper saves approximately 17 trees. \\
\cmidrule(lr){2-3}
&
(2) Energy Conservation and Lower Carbon Emissions &
 Recycling aluminum saves up to 95\% of the energy compared to producing new aluminum. \\
 \cmidrule(lr){2-3}

&
(3) Economic Benefits: Cost Savings and Job Creation &
Recycling creates about 10 jobs for every landfill job. \\
\cmidrule(lr){2-3}

&
(4) Reducing Pollution: Cleaner Air, Water, and Soil &
Unrecycled plastic breaks into microplastics that harm over 700 marine species and enter the food chain. \\
\cmidrule(lr){2-3}

&
(5) Setting a Positive Example and Meeting Social Responsibility Goals &
Companies with recycling programs report up to 20\% higher employee engagement and retention. \\

\midrule

\multirow{11}{1.2cm}{\textbf{Public transit}} &
(1) Environmental Benefits: Reduced Carbon Footprint &
Switching from personal cars to public transit can reduce an individual's transportation emissions by up to 50\%. \\
\cmidrule(lr){2-3}

&
(2) Cost Savings for Individuals and Businesses &
On average, commuters can save \$10{,}000+ per year by using public transit instead of driving. \\
\cmidrule(lr){2-3}

&
(3) Reduced Traffic Congestion &
A 10\% shift from personal cars to public transit can cut commute times by up to 40\%. \\
\cmidrule(lr){2-3}

&
(4) Health and Safety Improvements &
Traveling by public transit is approximately 10$\times$ safer than driving a personal car. \\
\cmidrule(lr){2-3}

&
(5) Economic and Social Benefits for Communities &
Every \$1 invested in public transit generates approximately \$4 in community benefits. \\

\bottomrule
\end{tabularx}
\caption{Fixed set of persuasive arguments and supporting facts used in the study for each topic.
All experimental modes used the same five argument--fact pairs for each topic, ensuring that participants across experimental modes had access to the same information. Each concrete argument is associated with a premise (e.g., ``Recycling uses less energy than making new materials'') which we omit here for brevity. See Appendix~\ref{appendix:key-arguments} for the complete arguments.} 
\label{tab:arguments-facts}
\vspace{-10pt}
\end{table*}

\subsection{Experiment Conditions}
\label{sec:study-design:exp-conditions}

The study employed a single-factor between-subjects design with three delivery mode corresponding to varying degrees of interactivity: essay, chat, and a text-based game.

\subsubsection{Essay}
\label{sec:study-design:essay}
Given the fixed set of arguments and facts, we asked GPT-4.1 to generate persuasive essays solely based on the facts and arguments (see Appendix~\ref{appendix:essay-generation-prompt} for the prompts). To ensure that all arguments and facts are covered in the essay, we inputted the essay and the arguments into the LLM to verify the coverage. We also randomly sampled essays and manually verified the coverage of arguments. Since the essay mode is not interactive, we opted for speeding up the computation time for generation and argument coverage verification by generating the essays in advance. However, to ensure the generality of the findings and for the sake of fair comparison to the interactive modes, rather than generating one essay per topic, we generated twenty essays and randomly sampled an essay for each participant.

\subsubsection{Chatbot}
\label{sec:study-design:chatbot}

In the chatbot condition, participants engaged in an interactive, free-form conversation with an LLM about the assigned sustainability topic. The chatbot presented the same fixed set of arguments and supporting facts used in the other conditions, but delivered them conversationally in response to participant inputs (see Table~\ref{tab:chat_template} for a chat example).

Unlike the essay condition, participants could actively steer the interaction by asking questions, requesting clarification, or reacting to the information presented, allowing them to control the pacing and order in which content was encountered.

To ensure comparability across conditions and enable later assessment of knowledge retention, the interaction was structured so that all predefined arguments and factual statements were eventually presented during the session. Participants could freely interact with the chatbot, and the conversation continued until all target facts had been covered or the participant chose to end the interaction. The interface indicated when all required content had been presented and allowed participants to conclude the session at that point. Appendix~\ref{appendix:chatbot-implementation} provides details about the content coverage verification mechanism.

\subsubsection{Text-based Game}
\label{sec:study-design:game}

In the text-based game condition, participants interacted with a narrative-driven, text-based game centered on the assigned sustainability topic (see Table~\ref{tab:game_template} for a game example). The game embedded the same fixed set of persuasive arguments and supporting facts used in the essay and chatbot conditions, but presented them within an interactive story world.

Participants assumed the role of a protagonist and progressed through the game by reading narrative descriptions and entering free-form text responses at each turn. The system responded by advancing the story, introducing new situations, characters, or dilemmas, while weaving in the predefined arguments and factual information in a contextualized manner.

As in the chatbot condition, participants could control the pacing and flow of the experience through their inputs. Similarly to the chatbot, the interface indicated when all required content had been covered and only then allowed participants to conclude the session, though participants could continue interacting until the ending of the game if they wished. Full implementation details of the text-based game, including plot generation and moderation prompts, are provided in Appendix~\ref{appendix:text-based-game}.

\begin{table}[ht]
    \centering
    \scriptsize
    \caption{Example conversation between \textcolor{cadmiumgreen}{player} and \textcolor{orange}{chatbot}. \textbf{Bolded lines} highlight factual information embedded into the chatbot's response.}
    \setlength{\tabcolsep}{4pt}
    \tt
    \begin{tabular}{|p{0.95\textwidth}|}
    \hline

        \textcolor{orange}{\textbf{Chatbot:}} Have you ever thought about how using public transit could make a real difference—for you and your community? Making the switch has more benefits than most people realize. Would you be open to hearing some of them?

         \textcolor{cadmiumgreen}{\textbf{Player:}} Not really, I would like to hear something new from you. \\
         \textcolor{orange}{\textbf{Chatbot:}} Absolutely, let’s start with something practical—money. \textbf{Did you know the average commuter saves over \$10,000 a year just by choosing public transit instead of driving?} That’s a lot of extra cash back in your pocket for things you’d actually enjoy. What would you do with those savings?

         \textcolor{cadmiumgreen}{\textbf{Player:}} Wow, that is a really shocking number, I think I should buy myself a new laptop and a new phone! \\
         \textcolor{orange}{\textbf{Chatbot:}}That’s an awesome way to use the extra money! Beyond the financial perks, there's another upside: \textbf{taking public transit actually cuts your transport emissions by 50\%.} It's a big, simple step toward a cleaner environment—and it’s as easy as hopping on a bus or train. Would you feel good knowing your daily commute could help clear up the air?\\         
         \textcolor{blue}{[The chat continues until all the facts are presented and the user chooses to end the session.]}\\   

        \hline
    \end{tabular}
    
    \label{tab:chat_template}
\end{table}

\begin{table}[ht]
    \centering
    \scriptsize
    \caption{Example conversation between \textcolor{cadmiumgreen}{player} and \textcolor{red}{game system}. \textbf{Bolded lines} highlight factual information seamlessly embedded into the story’s progression. At each turn, the player may choose one of three predefined actions (as in the first turn) or propose a custom response (as in the second turn).}
    \setlength{\tabcolsep}{4pt}
    \tt
    \begin{tabular}{|p{0.95\textwidth}|}
    \hline

        \textcolor{red}{\textbf{Game:}} Rain streaks across Sandbourne City Hall as you clutch the heavy “Sandbourne: 2040” folder outside the mayor’s office, your heartbeat echoing the drip outside. In the hallway, murmurs about record smog, traffic, and wet footprints braid into the city’s restless story. Your phone buzzes: six days until a decisive transportation vote. Data alone won’t save Sandbourne—you must turn numbers into a narrative. The door opens. “Alex? Ready to shape the future?” What do you do?
        \begin{enumerate}
            \item Enter confidently, ready to introduce bold ideas right away.
            \item Pause for a moment, listening to the worries in the hallway for clues about the public mood.
            \item Flip open your folder to review your talking points, wanting to be extra prepared before you go in.
            \item Or suggest your own move.
        \end{enumerate}

         \textcolor{cadmiumgreen}{\textbf{Player:}} 3 \\
         \textcolor{red}{\textbf{Game:}} You flip open your thick folder, eyes scanning cost projections and air quality charts. A bold, highlighted line catches your gaze: \textbf{“Switching to transit can cut your transport emissions by 50\%.”} Maya Alvarez, bright in her neon vest, rallies volunteers beneath a hazy skyline. “We need stories, not just science, Alex. Help us make people believe clean air is possible.” Nearby, Officer Don Briggs watches kids cough in the heat. Jared Liu leans on his car, smirking, “So, you turning us all into bus riders?” Between hope and skepticism, the city waits for your answer. What do you do?
        \begin{enumerate}
            \item Join Maya, helping plan her “Clean Air Now” event and share the 50\% emissions cut with the crowd.
            \item Engage Jared, linking pollution—and his own cough—to the transit facts you just reviewed.
            \item  Ask Officer Briggs for a ride-along to see firsthand how smog affects the community.
            \item Or suggest your own move.
        \end{enumerate}
         \textcolor{cadmiumgreen}{\textbf{Player:}} I help the volunteers to gather trash for a bit. \\
         \textcolor{red}{\textbf{Game:}} Gloves on, you join Maya picking up litter by the swings. Kids look grateful; Maya beams. “Actions speak, Alex,” she says. Nearby, Jared pauses mid-chuckle, watching you. Officer Briggs nods approval. As Maya helps you up, she whispers, “Help me get folks to the Clean Air rally tonight—and let them know \textbf{transit could cut our emissions in half.}” What do you do?\\...\\         \textcolor{blue}{[The game continues with options and interactions until all the facts are presented and the user chooses to end the session.]}\\   

        \hline
    \end{tabular}
    
    \label{tab:game_template}
\end{table}

\begin{figure*}[t]
    \centering
\includegraphics[width=\linewidth,trim={2cm 5cm 0cm 0cm},clip]{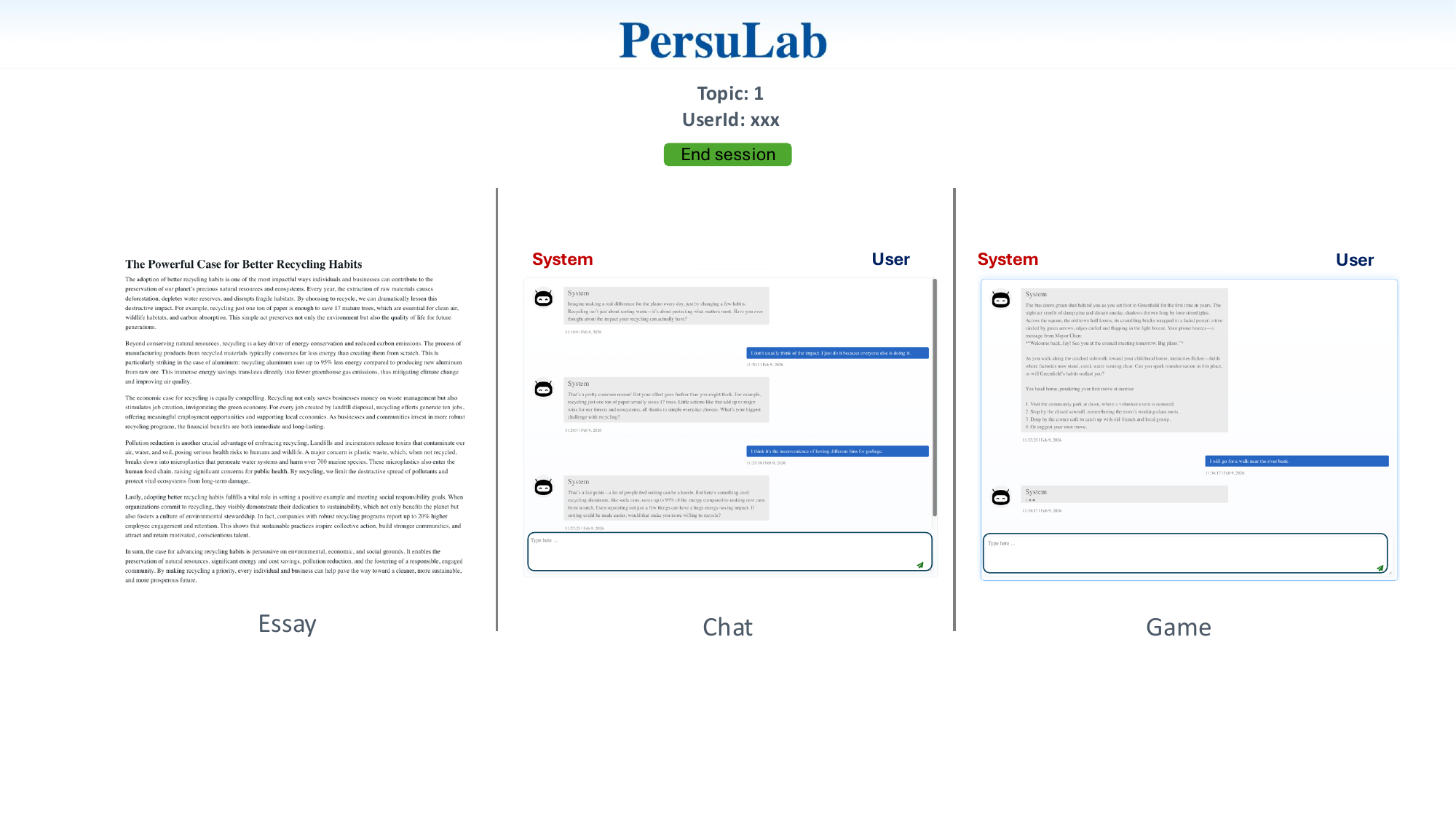}
    \caption{The user interface of PersuLab, illustrating the three system variants corresponding to different assignment conditions. Each interface displays a unique User ID assigned to the participant at the top. An “End Session” button becomes active once all facts have been presented in the chat and game conditions, and after three minutes of reading in the essay condition.}
 \label{fig_UI}
\end{figure*}

\subsection{\ourtool{}}
\label{sec:system}

We introduce \ourtool{}, a system designed to support both non-interactive and interactive information delivery while ensuring consistent exposure to a fixed set of persuasive arguments and factual statements.\footnote{The code is available at:\\\url{https://github.com/salavi/persulab}} Figure~\ref{fig_UI} shows the user interfaces across the three conditions. Upon entering the study, participants were assigned a topic and condition and were redirected to the corresponding interface. Participants were unaware of alternative conditions or topics. Each participant was assigned a unique user ID, enabling us to link between interaction logs and questionnaire responses while preserving anonymity. Across all conditions, an \textit{End Session} button was visible but activated only after condition-specific completion criteria were satisfied.

\begin{figure*}[t]
    \centering
\includegraphics[width=.9\textwidth,trim={0 0 0 0cm},clip]{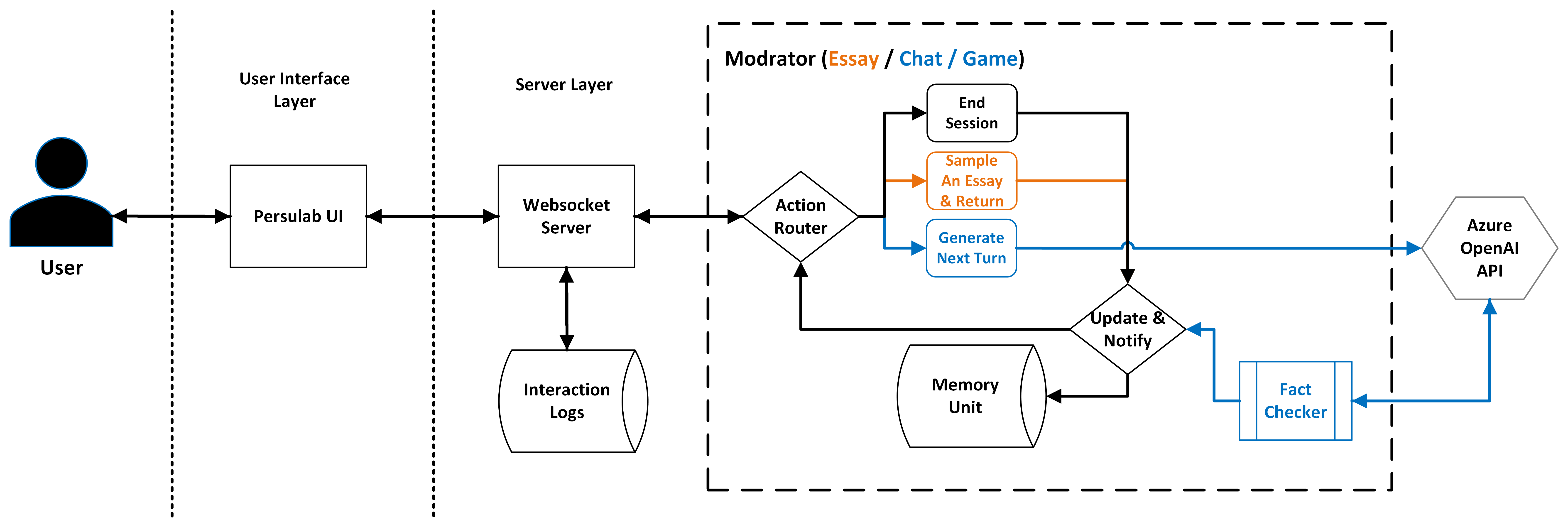}
    \caption{ System architecture and interaction workflow for PersuLab across experimental modes.
The \textcolor{orange}{essay condition (orange)} follows a single-pass content delivery pipeline using pre-generated essays that already cover all target facts. \textcolor{blue}{The chat and game conditions (blue)} operate through an interactive loop that includes turn generation, fact coverage tracking, and memory updates. Black paths denote shared system components and data flow common to all modes (e.g., session control and logging). All interactions are logged for subsequent analysis.}
 \label{fig_system_workflow}
\end{figure*}
Figure~\ref{fig_system_workflow} presents an overview of \ourtool{}'s system architecture and content generation workflows across the three experimental conditions. In the essay condition, content delivery follows a single-pass pipeline. Because factual coverage is guaranteed during pre-generation (see Section~\ref{sec:study-design:essay}), no runtime fact tracking or verification is required. The minimum exposure period of 60 seconds was introduced to ensure participants had sufficient time to read the essay before proceeding.

In contrast, the chatbot and text-based game conditions follow an interactive generation loop. Participant inputs are routed to a moderator module that constructs prompts and invokes the language model to generate the next system response. After each generated turn, an automated fact-checking module evaluates which of the remaining target facts have been covered and updates the system's internal record of uncovered content. This updated information is incorporated into subsequent prompts to guide future generations and ensure complete coverage of all predefined facts. In our experiment, after all predefined facts had been presented, participants were free to end the session or continue interacting until either the interaction naturally concluded or the maximum interaction time of 25 minutes was reached.

Both interactive conditions maintain a shared session state that includes interaction history, coverage status of target facts, and timing metadata. All interactions across conditions are logged for analysis, including turn-level content, timestamps, and fact coverage information. Detailed prompt structures, generation parameters, and fact-checking mechanisms for each condition are provided in Appendices~\ref{appendix:essay-generation-prompt}, \ref{appendix:chatbot-implementation}, and \ref{appendix:text-based-game}, respectively. 

\subsection{Participants}
\label{sec:study:participants}

We recruited 45 participants through university advertisements and word of mouth. Of these, 43 participants were included in the analysis; one participant did not complete all study steps, and one participated in a dry run. Most participants were between 18–34 years old (Fig.~\ref{fig:age-dist}). Participants were approximately evenly distributed across the three experimental conditions (essay, chat, and game) and across the two topics (recycling and public transit) as shown in Fig. ~\ref{fig:assign-dist}.

\begin{figure}[t]
    \centering


    \begin{subfigure}{0.5\linewidth}
        \centering
        \scriptsize
        \begin{tikzpicture}
        \begin{axis}[
            xbar,
            width=\linewidth,
            height=3cm,
            xmin=0,
            xmax=30, 
            xlabel={Number of participants},
            ytick=data,
            yticklabels={18--24,25--34,35--44,45--54},
            y dir=reverse,
            bar width=7pt,
            axis x line*=bottom,
            axis y line*=left,
            ticklabel style={font=\scriptsize},
            label style={font=\scriptsize},
            nodes near coords,
            nodes near coords align={horizontal},
            every node near coord/.append style={font=\scriptsize, xshift=2pt},
        ]
        \addplot coordinates {(28,0) (11,1) (3,2) (1,3)};
        \end{axis}
        
        \end{tikzpicture}
        \caption{Age-group distribution (n=43).}
        \label{fig:age-dist}
    \end{subfigure}
    \hfill
    \begin{subfigure}{0.47\linewidth}
    
        \centering
        \scriptsize
        \begin{tabular}{lcccc}
            \toprule
             & Chat & Essay & Game & Total \\
            \midrule
            Recycling & 8 & 7 & 7 & 22 \\
            Public transit & 6 & 7 & 8 & 21 \\
            \midrule
            Total & 14 & 14 & 15 & 43 \\
            \bottomrule
        \end{tabular}
        \vspace{15pt}
        \caption{Assignment counts by mode and topic}
        \label{fig:assign-dist}
    \end{subfigure}
\caption{Participant demographics and assignment distribution.}
\label{fig:participants-summary}
\end{figure}

\section{Evaluation Measures}
\label{sec:evaluation-measures}

We evaluate the knowledge retention and persuasive effectiveness of the three experiment modes through a mixed-methods user study. We compare participants' attitudes towards the assigned topic collected before the study (\S\ref{sec:evaluation-measures:pre-study-baselines}) with their \textbf{subjective experience and perceived attitude changes} after the study (\S\ref{sec:evaluation-measures:subjective-exp}). To measure objective knowledge retention, participants had to complete a delayed \textbf{knowledge assessment task} (\S\ref{sec:evaluation-measures:objective-knowledge-scores}). Finally, we analyze the \textbf{interaction logs}  to probe potential relationships between interaction dynamics and subjective and objective learning outcomes (\S\ref{sec:evaluation-measures:interaction-metrics}). 

\subsection{Pre-Study Questionnaire}
\label{sec:evaluation-measures:pre-study-baselines}

Prior to the experimental session, participants completed a pre-study questionnaire capturing baseline attitudes and behaviors relevant to the assigned topic. These measures were used to characterize participants' initial positions. We collected baseline measures of \textbf{importance}, \textbf{behavioral intention}, and \textbf{epistemic confidence}, which are detailed below (see Appendix~\ref{appendix:questionnaires:pre-study} for the complete pre-study questionnaire). 

Participants were asked to rate the \textbf{importance} of the topic to them on a 5-point Likert scale. For \textbf{behavioral intention}, we inquired about existing habits pertaining to the topic, such as frequency of recycling or commuting by public transport. Responses were recoded into ordinal scales (1 $\rightarrow$ 5) to reflect increasing levels of baseline engagement (e.g., never recycle to always recycle). These measures provide descriptive context about participants' pre-study attitudes and behaviors.

We also collected a pre-study measure of participants' \textbf{self-perceived confidence in their knowledge of the topic} using a 5-point Likert scale. We treat \textit{epistemic confidence} as a descriptive indicator of participants' perceived familiarity with the topic, which provides context for interpreting objective knowledge retention and subjective learning outcomes. 

Additional demographic and contextual questions (e.g., age, access to recycling services, access to a personal car) were collected for descriptive purposes but were not used in the primary analyses. 

\subsection{Post-Study Questionnaire}
\label{sec:evaluation-measures:subjective-exp}

Immediately after completing the intervention, participants rated their subjective experience using a set of 5-point Likert-scale items (1 = strongly disagree, 5 = strongly agree). We included questions directly targeting learning and persuasion outcomes as well as questions about subjective experience common in user studies, as we detail below. See  Appendix~\ref{appendix:questionnaires:post-study} for the complete set of questions. 

\paragraph{Perceived Change and Persuasion.} To assess persuasive impact, participants reported \textbf{perceived changes} in their attitudes relative to before the session. For each topic, participants indicated whether the interaction led to a change in \textbf{importance}, \textbf{behavioral intention}, and \textbf{belief in effectiveness}, using ordered categorical responses (Less, Same, More, Not sure). Participants additionally rated the convincingness of the arguments presented using 5-point Likert scales. Finally, we prompted participants to describe their qualitative reflections in free-text format.

\paragraph{Subjective Experience.} We assessed participants' perceived \textbf{ease of following} the provided content, \textbf{engagement}, \textbf{enjoyment}, \textbf{trust} in the information, \textbf{self-reported learning}, \textbf{satisfaction}, \textbf{convincingness of arguments}, 
\textbf{increased motivation to act}, overall \textbf{influence on thinking}, as well as their willingness to \textbf{recommend others} and \textbf{re-encounter in future} similar experiences using a set of 5-point Likert-scale items. 

\subsection{Delayed Objective Knowledge Assessment}
\label{sec:evaluation-measures:objective-knowledge-scores}
To measure objective knowledge retention, we reached out to participants \textbf{24 hours after the experiment} and asked them to complete a multiple-choice knowledge quiz based on the facts covered during the experiment. The 24-hour delay is motivated by prior research in psychology and HCI that suggests that an overnight consolidation period facilitates measuring knowledge retention rather than immediate knowledge recall \cite{cepeda2006distributed, roediger2006test, kantak2012learning, anderson2013gesture}.

For each topic, the quiz included five content-covered questions and two control questions referencing information not provided in the experience. Control questions were included to discourage guessing or external lookup; selecting ``I have not seen this information before'' was treated as the correct response for these items.

Participants' knowledge scores were computed only as the number of correct responses to the five content-covered questions. Confidence ratings were collected after each question but were not analyzed in this work. The full set of knowledge quiz items and correct answers for each topic is reported in Appendix~\ref{appendix:questionnaires:knowledge-quiz}.

\subsection{Interaction Log Metrics (Exploratory)}
\label{sec:evaluation-measures:interaction-metrics}

For the interactive conditions (chat and game), we additionally recorded participants' \textbf{interaction logs}. These included the number of user/system turns, total and mean user/system words per turn, user-to-system word ratios, session duration, and mean user reaction time (time between received system message and user's next response).

Because these measures were not pre-registered and involved multiple comparisons, analyses of interaction logs is treated as \textbf{exploratory}. We use them to probe potential relationships between interaction dynamics and subjective experience or persuasive outcomes, rather than to test confirmatory hypotheses. Accordingly, findings from these analyses are reported descriptively and interpreted with appropriate caution.
\section{Analysis Methods}
\label{sec:analysis-methods}

Our analyses are designed to examine the following aspects. First, we look at
differences between experimental modes in subjective experience, answering the question ``which experimental mode do participants prefer?'' (\S\ref{sec:analysis:subjective}). Second, we examine the perceived changes in persuasive constructs (e.g., importance, behavioral intention, and belief in effectiveness), essentially answering the question ``which experimental mode do participants perceive as more persuasive?'' (\S\ref{sec:analysis:persuasive_constructs}). We then report the results of the objective knowledge retention test, answering the question ``which experimental mode is the most effective for teaching participants about the topic?'' 
(\S\ref{sec:analysis:objective}). Finally, we perform an exploratory analysis of the relationships between properties of the participant-system interactions (e.g., user word counts) and persuasive outcomes (\S\ref{sec:analysis:exploratory}).

Throughout the analysis, we report descriptive statistics alongside inferential results to support transparent interpretation. All significance tests are clearly labeled as confirmatory or exploratory, and findings are interpreted accordingly.

\subsection{Subjective Experience}
\label{sec:analysis:subjective}

To compare post-study subjective experience ratings (e.g. ease of following, engagement, etc.) across the three experimental modes (essay, chat, game), we used nonparametric tests appropriate for \textbf{ordinal} Likert-scale data.

Specifically, we first applied \textbf{Kruskal--Wallis} \cite{kruskal1952use} tests to assess overall differences across the three modes. When a significant omnibus effect was observed, we conducted pairwise comparisons using \textbf{Mann--Whitney U} \cite{mann1947test} tests (essay vs game, essay vs chat, chat vs game). These tests were chosen because Likert-scale responses are ordinal and may violate the assumptions of normality required for parametric tests.

All tests were two-sided, and statistical significance was assessed at $\alpha = .05$. Effect directions and descriptive statistics (means and standard deviations) are reported alongside $p$-values to aid interpretation.

\subsection{Perceived Change and Persuasion}
\label{sec:analysis:persuasive_constructs}

To analyze perceived changes in importance, behavioral intention, and belief in effectiveness, we modeled participants' self-reported change using \textbf{ordinal logistic regression} \cite{mccullagh1980regression}. Perceived change responses (Less / Same / More) were encoded as ordinal change scores ($-1$, $0$, $+1$), with ``Not sure'' responses treated as missing. Separate models were fit for each construct (importance, intention, belief), both within-topic (recycling-only and transit-only) and using a combined-topic dataset in which equivalent constructs were aligned across topics.

For each outcome, we first fit an omnibus three-condition model of the form:
\[
\text{Change}_i \sim \text{Mode}_i 
\]

where $\text{Change}_i$ is the ordinal change score for participant $i$,
$\text{Mode}_i$ is the experimental condition (essay as the baseline, with dummy variables for chat and game).

To further probe differences between specific modes, we additionally conducted pairwise ordinal logistic regressions between each pair of conditions 
using the same model specification but restricted to the relevant subset of data.

All ordinal logistic regression models (ordered logit / proportional-odds) were estimated via maximum likelihood using a logit link function (BFGS optimizer). Predictor coefficients are reported on the log-odds scale, and statistical significance was assessed using Wald $z$-tests \cite{wald1943tests}.

\paragraph{\textbf{Baseline Robustness Checks.}}
For importance and behavioral intention, conceptually aligned pre-study measures were available. As a robustness check, we re-ran the analyses including the corresponding pre-study scores as covariates to account for potential ceiling effects. The inclusion of baseline measures did not qualitatively change the pattern of results or the relative differences between modes (see Appendix~\ref{appendix:baseline-effects}).

\subsection{Objective Knowledge Retention Analyses}
\label{sec:analysis:objective}

To compare the participants' knowledge scores across experimental modes, we again used Kruskal--Wallis tests followed by Mann--Whitney U tests for pairwise comparisons, due to the ordinal and bounded nature of the score distribution and the modest sample size. Control questions and confidence ratings were excluded from the score computation and analyses.

\subsection{Exploratory Interaction Log Analyses}
\label{sec:analysis:exploratory}

For the interactive conditions (chat and game), we explored relationships between interaction log metrics (\S\ref{sec:evaluation-measures:interaction-metrics}) and outcome measures: subjective experiences (\S\ref{sec:evaluation-measures:subjective-exp}) and objective knowledge scores (\S\ref{sec:evaluation-measures:objective-knowledge-scores}), using Spearman rank-order correlations \cite{Zar2005}.

Because these analyses involved multiple comparisons and were not pre-registered, they were treated as exploratory. No formal correction for multiple comparisons was applied; instead, results are reported descriptively, with emphasis on effect direction and consistency rather than statistical significance alone. These analyses are intended to generate hypotheses for future work rather than to support confirmatory claims.

\section{Findings}
\label{sec:findings}

Unless otherwise noted, we report results pooled across the two topics (recycling and public transit) and compare outcomes by delivery mode. This decision was motivated by the study design: splitting the data by both topic and mode would yield small cell sizes (approximately 6--7 participants per condition), limiting interpretability and statistical stability.

Additionally, pooling across topics is supported by the observation that topic-specific ratings were highly similar across key measures. In particular, perceived convincingness of arguments was comparable between recycling (mean $=3.82$) and public transit (mean $=3.76$), suggesting that topic differences did not meaningfully influence participants' evaluations. Accordingly, we focus on mode-level comparisons throughout this section.

This section is structured to mirror Sec.~\ref{sec:analysis-methods}. We report the participants' subjective experience with the various experimental modes (\S\ref{sec:findings:subjective-outcomes}); the perceived changes in persuasive constructs (\S\ref{sec:findings:pre-post-effects}); the results of the objective knowledge test (\S\ref{sec:findings:objective-knowledge}); and finally an exploratory analysis of the relationships between properties of the participant-system interactions and persuasive outcomes (\S\ref{sec:findings:log-analysis}).

\subsection{Subjective Experience}
\label{sec:findings:subjective-outcomes}

\begin{figure*}[ht]
    \centering
\includegraphics[width=\textwidth,trim={0 0 0 0cm},clip]{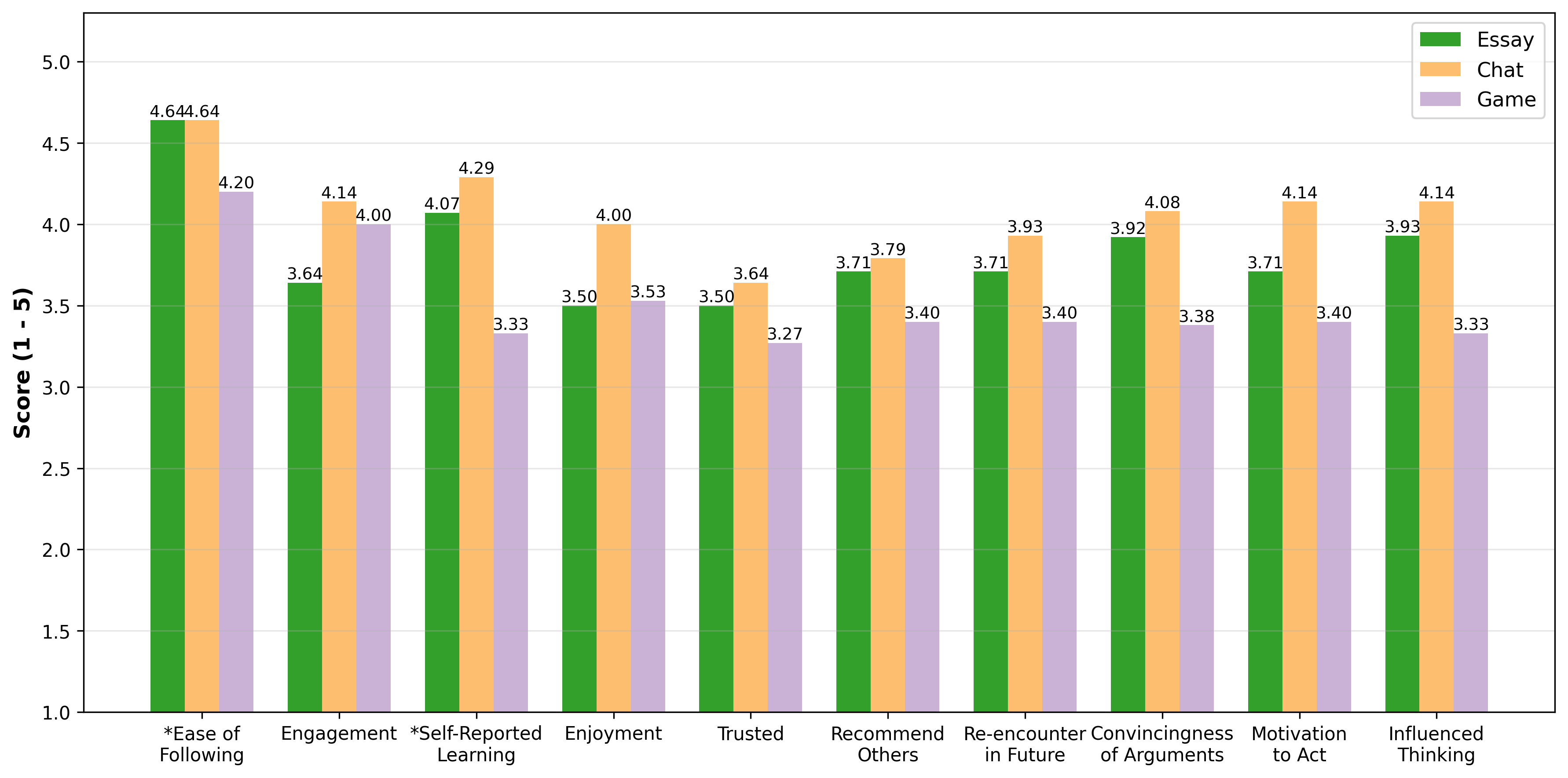}
    \caption{Mean subjective outcome ratings (5-point Likert scale) across delivery modes, aggregated across topics (n = 43). Asterisks (*) denote measures with statistically significant differences across conditions.}
 \label{subjective_overall_chart}
\end{figure*}

Participants' subjective experiences across experimental conditions are summarized in Figure~\ref{subjective_overall_chart}. Overall, subjective ratings were relatively high across all three conditions, with mean scores generally above the midpoint of the scale (3) for most measures. Across dimensions, the chatbot condition was consistently rated highest, while the game condition tended to receive lower ratings. The essay condition generally fell between the two interactive conditions, depending on the measure.

\paragraph{\textbf{Ease of following.}}
The essay and chatbot conditions received the highest ease-of-following ratings (both Mean $= 4.64$), whereas the game condition was rated lower (Mean $= 4.20$). A Kruskal–Wallis test revealed a significant effect of delivery mode on ease of following ($p = 0.0497$). Post-hoc Mann–Whitney U tests indicated that participants found the essay condition significantly easier to follow than the game condition ($p = 0.03$). Additionally, the difference between the chatbot and game conditions approached statistical significance ($p = 0.07$). It is possible that the additional cognitive demands introduced by narrative structure, role-based interaction, or the verbosity of the system in the game mode contributed to this finding.

\paragraph{\textbf{Self-reported learning.}}
Participants were asked to rate how much they believed they learned during the experiment. Ratings were highest for the chatbot condition (mean $= 4.29$), followed by the essay condition (mean $= 4.07$), and lowest for the game condition (mean $= 3.30$). A Kruskal–Wallis test indicated a significant effect of delivery mode on self-reported learning ($p = 0.018$). Post-hoc pairwise Mann–Whitney tests revealed that participants reported significantly higher learning in the chatbot condition compared to the game condition ($p = 0.005$). The difference between the essay and chatbot conditions approached significance ($p = 0.055$).

\paragraph{\textbf{Engagement.}} 
As expected, given the interactive nature of the conditions, engagement ratings were higher for the chatbot (Mean $=4.14$) and game (Mean $=4.00$) compared to the essay condition (Mean $=3.64$), suggesting that interactivity may enhance participants' sense of involvement relative to passive content consumption. Among the three modes, the chatbot received the highest engagement ratings overall, consistent with more extensive user contributions observed in the chatbot condition (\S~\ref{sec:findings:log-analysis})

\paragraph{\textbf{Enjoyment.}}
Interestingly, enjoyment ratings showed a more modest separation across conditions. While the chatbot condition received the highest mean enjoyment rating (Mean $=4.00$), the game (Mean $=3.53$) and essay (Mean $=3.50$) conditions were rated similarly. One possible explanation for the slightly lower enjoyment ratings in the game condition in comparison to the chat is its text-heavy format and lack of visual elements.

\paragraph{\textbf{Trust and Perceived Persuasion}}
Measures related to perceived persuasion and behavioral impact -- including convincingness of arguments, motivation to act, and influence on thinking -- showed a consistent numerical advantage for the chatbot condition (means of $4.08$, $4.14$, and $4.14$, respectively). The essay condition followed closely (means of $3.92$, $3.71$, and $3.93$), while the game condition received lower ratings on average (means of $3.38$, $3.40$, and $3.33$). A similar pattern was observed for trust, with participants reporting higher trust in the chatbot condition (mean $= 3.64$) than in the essay (mean $= 3.50$) and game (mean $= 3.27$) conditions.

\paragraph{\textbf{Recommendation and Future Re-encounter.}} Following all the above ratings, participants rated the chatbot condition higher in terms of their willingness to recommend the experience to others and to re-encounter similar experiences in the future (means of $3.79$ and $3.93$, respectively), compared to the essay condition (means of $3.71$ and $3.71$). The game condition received the lowest ratings on both measures (means of $3.40$ and $3.40$).

\subsection{Perceived Change and Persuasion}
\label{sec:findings:pre-post-effects}

We examined perceived changes in participants' attitudes toward the target behaviors using post-study self-reports of \textit{perceived change} across three constructs: \textit{importance}, \textit{intention to act}, and \textit{belief in effectiveness}. These items explicitly asked participants to assess how their attitudes had changed relative to before the session (Less / Same / More), and were analyzed directly as ordinal outcomes.

\begin{figure*}[t]
    \centering
\includegraphics[width=\textwidth,trim={0 0 0 0cm},clip]{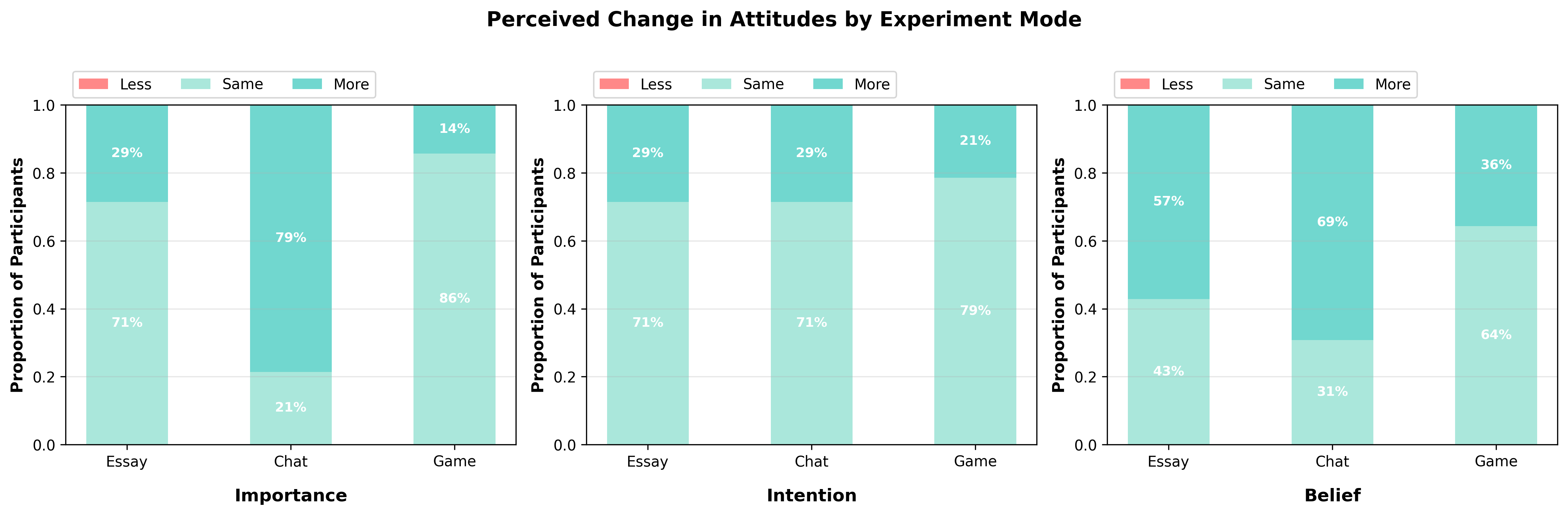}
    \caption{Stacked bar chart showing the proportion of participants experiencing perceived change across three constructs: Importance, Intention, and Belief. Each bar represents one of three experiment modes (Essay, Chat, Game), with color-coded segments. Percentages within each segment show the proportion of participants in that category.}
 \label{fig:perceived-change}
\end{figure*}

Figure~\ref{fig:perceived-change} shows the distribution of perceived change responses across experimental conditions; no participant reported a decrease. Descriptively, participants in the chatbot condition more frequently reported increases in perceived importance (79\%) and belief in effectiveness (69\%) than those in the essay (29\% and 57\%) and game (14\% and 36\%) conditions. In contrast, perceived changes in intention to act were more evenly distributed across conditions (29\% chatbot, 29\% essay, and 21\% game).

To formally assess mode effects we fit ordinal logistic regression models predicting perceived change scores from delivery mode (see Sec \ref{sec:analysis:persuasive_constructs}). In the combined-topic model, delivery mode had a significant effect on perceived change in importance. Participants in the chatbot condition reported significantly greater increases in importance than those in the essay condition ($\beta = 2.22$, $p = 0.012$),
while the game condition did not differ significantly from the essay ($\beta = -0.88$, $p = 0.37$). Pairwise models confirmed that chatbot interactions led to larger perceived increases in importance than both essay ($\beta = 2.22$, $p = 0.012<0.05$)
and game ($\beta = -3.09$, $p = 0.002<0.05$).

No significant mode effects were observed for perceived changes in intention or belief in effectiveness. 

These results indicate that delivery mode primarily influenced participants' perceptions of issue importance, rather than directly shaping beliefs about effectiveness or stated behavioral intentions. A baseline robustness analysis incorporating pre-study measures of importance and behavioral intention yielded consistent results (Appendix~\ref{appendix:baseline-effects}), suggesting that these effects are not driven by pre-existing differences between participants.

One interpretation of this pattern is that recognizing a topic as effective does not necessarily translate into perceiving the issue as personally important, nor does either guarantee changes in intended behavior. The findings in this section therefore highlight a distinction between \textit{belief updating}, \textit{issue salience}, and \textit{behavioral intention}. While providing information appears sufficient to strengthen beliefs about effectiveness across modes, making an issue feel personally important is more sensitive to interaction format.

At the same time, changes in behavioral intention were limited, likely reflecting both the inherent difficulty of motivating action and participants' high baseline engagement with the target behaviors. Indeed, pre-study measures show that most participants already reported frequent recycling or regular public transit use (Appendix~\ref{appendix:prestudy-baselines}), leaving limited room for further increases in stated intention.

Overall, these results indicate that conversational interaction can meaningfully increase the perceived importance of sustainability-related behaviors beyond static text, even when informational content is held constant.

\subsection{Objective Knowledge Retention}
\label{sec:findings:objective-knowledge}

\begin{figure*}[t]
    \centering
\includegraphics[width=\textwidth,trim={0 0 0 0cm},clip]{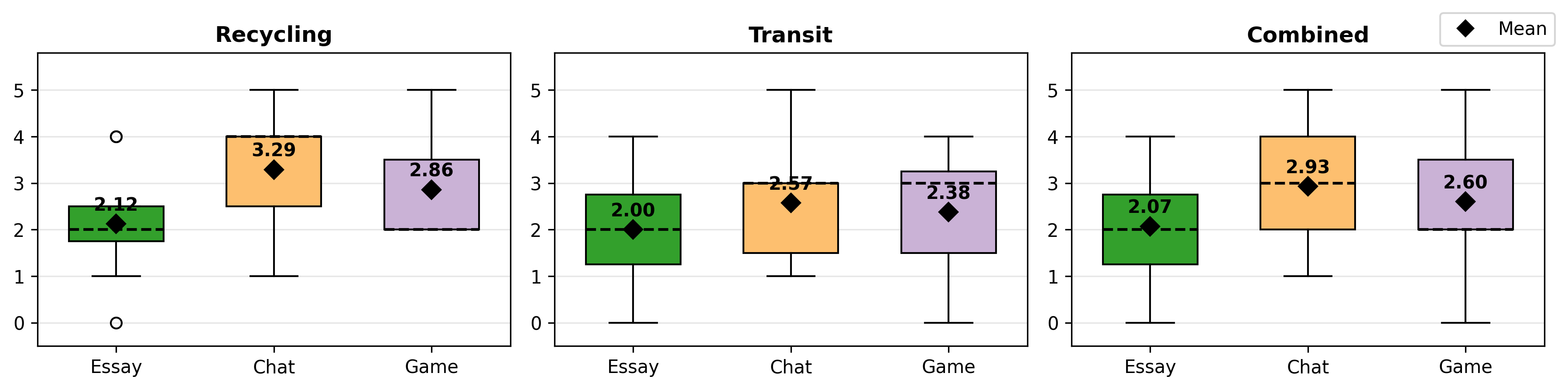}
    \caption{Delayed (24-hour) objective knowledge retention scores by delivery mode (essay, chatbot, text-based game), shown separately for the recycling topic, public transit topic, and the combined dataset. Scores indicate the number of correctly answered content-covered questions (out of five). Diamonds denote mean scores; boxplots show the distribution across participants.
}
 \label{objective-knowledge-scores}
\end{figure*}

Figure~\ref{objective-knowledge-scores} shows participants' performance on the delayed (24-hour) objective knowledge quiz, reported separately for each topic and for the combined dataset. Scores reflect the number of correctly answered content-covered questions (out of five).

Across both topics and in the combined analysis, participants in the \textbf{chatbot} condition achieved the highest mean knowledge scores, followed by the \textbf{text-based game}, with the \textbf{essay} condition scoring lowest on average. In the combined dataset, mean scores were $M=2.93$ for chatbot, $M=2.60$ for game, and $M=2.07$ for essay.

Surprisingly, participants in the game condition consistently outperformed those in the essay condition despite reporting lower self-assessed learning in the post-study questionnaire (\S\ref{sec:findings:subjective-outcomes}). This pattern was observed for both recycling and public transit topics, as well as in the pooled analysis. However, variability within conditions was substantial, and differences across modes did not reach statistical significance.

These results suggest that interactive modes may support stronger factual retention over time, even when participants do not explicitly report greater learning.

\subsection{Exploratory Interaction Log Analyses}
\label{sec:findings:log-analysis}

As shown in Table~\ref{tab:interaction-logs-summary}, the game condition involved substantially more back-and-forth interaction, with higher numbers of both user and system turns and longer session durations. However, user contributions in the game condition were considerably shorter, while system responses were longer and more verbose, resulting in a much lower user–system word ratio compared to the chat condition. Participants in the game condition also responded more quickly on average. This pattern reflects the game’s interaction structure, in which many turns involved system-posed multiple-choice prompts. Although participants could suggest their own actions, they more often selected one of the predefined options, resulting in shorter user responses. 

\begin{table*}[t]
\centering
\footnotesize
\setlength{\tabcolsep}{6pt}
\renewcommand{\arraystretch}{1.15}
\begin{tabularx}{0.5\textwidth}{p{4.2cm} rr}
\toprule
\textbf{Metric} & \multicolumn{1}{c}{\textbf{Chat}} & \multicolumn{1}{c}{\textbf{Game}} \\
\midrule

Avg. User Turns              & 7.25   & 13.62  \\
User Total Words             & 150.50 & 40.54  \\
User Mean Words / Turn       & 22.45  & 2.92   \\
System Total Words           & 533.92 & 1892.31 \\
System Mean Words / Turn     & 63.82  & 131.05 \\
User--System Word Ratio      & 0.303  & 0.020  \\
Session Duration (sec)       & 576.35 & 783.12 \\
Mean Reaction Time (sec)     & 67.93  & 48.62  \\

\bottomrule
\end{tabularx}
\caption{Interaction log statistics by experimental condition.
The table summarizes turn-taking behavior, message length, temporal characteristics, and relative user--system contribution across chat and game modes.}
\label{tab:interaction-logs-summary}
\end{table*}

To examine whether interaction patterns were associated with subjective and objective outcomes, we conducted exploratory Spearman rank correlations between interaction metrics and outcome variables. Correlations were computed both across combined chat and game participants (\S\ref{sec:findings:exploratory:combined}) and separately within each condition (\S\ref{sec:findings:exploratory:separate}). See Sec~\ref{sec:analysis:exploratory} for the experimental setup.  

Several associations reached nominal significance (uncorrected), particularly within the combined and chat conditions. However, after controlling for multiple comparisons using the Benjamini–Hochberg false discovery rate (FDR) procedure across all tests, none of the correlations remained statistically significant.

We therefore interpret these associations cautiously as exploratory trends rather than confirmatory evidence of systematic relationships.


\begin{table*}[h]
\centering
\footnotesize
\setlength{\tabcolsep}{2pt}
\renewcommand{\arraystretch}{1.25}
\begin{tabularx}{\textwidth}{p{1.25cm} p{2.75cm} c c X}
\toprule
\textbf{Outcome Variable} & \textbf{Interaction Metric} & \textbf{$\rho$} & \textbf{$p$} & \textbf{Interpretation (Descriptive)} \\
\midrule

\multirow{7}{1.25cm}{Self-Reported Learning} & User turns & -0.396 & 0.0498 & More user turns are associated with lower perceived learning. \\

& User total words & +0.477 & 0.0159 & Higher user word count is associated with higher perceived learning. \\

& User avg. words/turn & +0.501 & 0.0107 & Longer user messages are associated with higher perceived learning. \\

& System total words & -0.530 & 0.0065 & Greater system verbosity is associated with lower perceived learning. \\

& System avg. words/turn & -0.444 & 0.0262 & Longer system messages are associated with lower perceived learning. \\

& User--system word ratio & +0.519 & 0.0079 & Higher relative user contribution is associated with higher perceived learning. \\

& Session duration & -0.439 & 0.0279 & Longer sessions are associated with lower perceived learning. \\ \midrule

Trust & Mean reaction time & +0.399 & 0.0481 & Longer response latencies are associated with higher trust ratings. \\

\bottomrule
\end{tabularx}
\caption{Exploratory Spearman correlations between interaction metrics and outcome variables for combined chat and game participants. After applying the Benjamini--Hochberg false discovery rate (FDR)~\cite{benjamini1995controlling} correction across all tested correlations, none of the associations remained statistically significant. Only correlations with raw $p$-value $<0.05$ are reported in this table.}
\label{tab:spearman-combined}
\end{table*}

\subsubsection{\textbf{Combined game and chat logs.}}
\label{sec:findings:exploratory:combined}

As shown in Table~\ref{tab:spearman-combined}, several interaction metrics exhibited nominal associations with self-reported knowledge when considering combined participants. Greater \textbf{user contribution} (reflected in higher total user word counts ($\rho = .477$, $p = .016$), longer average user messages ($\rho = .501$, $p = .011$), and a higher user--system word ratio ($\rho = .519$, $p = .008$)) was positively associated with perceived learning. In contrast, higher \textbf{system verbosity} (system total words: $\rho = -.530$, $p = .007$; system mean words per turn: $\rho = -.444$, $p = .026$) was \textbf{negatively} associated with \textbf{self-reported learning}. These findings may suggest that users felt they were learning more from more interactive settings in which they were engaged. 

Other factors that were \textbf{negatively} associated with \textbf{perceived knowledge} were a greater \textbf{number of user turns} ($\rho = -.396$, $p = .050$) and \textbf{longer session durations} ($\rho = -.439$, $p = .028$). These patterns mirror the structural differences observed between conditions in Table~\ref{tab:interaction-logs-summary}, where the game condition involved more turns, longer sessions, and more verbose system responses, while the chat condition involved longer user messages and higher relative user contribution.

Finally, mean reaction time showed a modest positive association with trust ($\rho = .399$, $p = .048$). Notably, response latencies were longer on average in the chat condition compared to the game condition (67.93 vs. 48.62 seconds), consistent with the higher trust ratings observed for chat in Section~\ref{sec:findings:subjective-outcomes}.

We ran Spearman rank correlations between mean reaction time and interaction metrics to disentangle whether longer response latencies reflected increased reading load or greater user contribution. Mean reaction time was positively associated with mean user message length ($\rho = .47$, $p = .019$) and the user–system word ratio ($\rho = .45$, $p = .025$), but not with system verbosity (trend only, $p \approx .08$). This suggests that longer response time -- particularly in the chat condition -- primarily reflect increased effort in composing responses rather than processing longer system outputs.


\begin{table*}[t]
\centering
\scriptsize
\setlength{\tabcolsep}{2pt}
\renewcommand{\arraystretch}{1.25}
\begin{tabularx}{\textwidth}{p{1.3cm} p{2.75cm} c c X}
\toprule
\textbf{Outcome Variable} & \textbf{Interaction Metric} & \textbf{$\rho$} & \textbf{$p$} & \textbf{Interpretation (Descriptive)} \\
\midrule

\multirow{2}{1.3cm}{Easy to follow} 
& User avg. words/turn 
& -0.717 
& 0.0087 
& Longer user messages are associated with lower ease of following the experience. \\

& User--system word ratio 
& -0.615 
& 0.0335 
& Higher relative user contribution is associated with lower convenience ratings. \\ 
\midrule

Engagement 
& User avg. words/turn 
& -0.656 
& 0.0205 
& Longer user messages are associated with lower reported engagement. \\
\midrule

Enjoyment 
& User avg. words/turn 
& -0.700 
& 0.0113 
& Longer user messages are associated with lower satisfaction. \\
\midrule

Encounter Again 
& User total words 
& -0.683 
& 0.0144 
& Higher user word count is associated with lower desire to repeat the experience. \\
\midrule

\multirow{3}{1.3cm}{Motivation to Act} 
& User turns 
& +0.598 
& 0.0402 
& More user turns are associated with higher motivation to act. \\

& User avg. words/turn 
& -0.702 
& 0.0110 
& Longer user messages are associated with lower motivation to act. \\

& User--system word ratio 
& -0.670 
& 0.0172 
& Higher relative user contribution is associated with lower motivation to act. \\

\bottomrule
\end{tabularx}
\caption{Exploratory Spearman correlations between interaction metrics and subjective outcomes for the chat condition. Raw $p$-values are reported. After applying the Benjamini--Hochberg false discovery rate (FDR) correction across all tested correlations, none of the associations remained statistically significant.}
\label{tab:spearman-chat}
\end{table*}

\subsubsection{\textbf{Within chat and game logs.}}
\label{sec:findings:exploratory:separate}

When examining correlations separately within each condition, nominal associations were observed primarily in the chat condition (Table~\ref{tab:spearman-chat}). Longer user messages (user avg. words/turn) were negatively associated with several subjective outcomes, including convenience ($\rho = -.717$, $p = .009$), engagement ($\rho = -.656$, $p = .021$), satisfaction ($\rho = -.700$, $p = .011$), and motivation to act ($\rho = -.702$, $p = .011$). Similarly, a higher user--system word ratio was negatively associated with convenience ($\rho = -.615$, $p = .034$) and motivation to act ($\rho = -.670$, $p = .017$). Greater total user word count was also negatively associated with the desire to encounter the experience again ($\rho = -.683$, $p = .014$).

In contrast, a greater number of user turns was positively associated with motivation to act ($\rho = .598$, $p = .040$), suggesting that more back-and-forth interaction aligned with higher reported motivation in the chat condition.

No nominally significant associations were observed within the game condition. As with the combined analysis, none of the correlations within either condition remained statistically significant after applying the Benjamini--Hochberg FDR correction. These results are therefore interpreted as exploratory trends rather than confirmatory evidence of systematic within-condition relationships.
\section{Discussion}
\label{sec:discussion}

Our goal in this work was to examine how three different modes of information delivery (essay, chatbot, and text-based game) shape persuasive experience and learning outcomes when informational content is held constant. By controlling for arguments and facts across conditions, our study isolates the role of interaction format and presentation style from content differences. The findings reveal several systematic tensions: interactive formats can enhance engagement and perceived importance without uniformly increasing behavioral intention; subjective learning judgments can diverge from objective knowledge retention; and narrative game-based experiences may support memory while simultaneously raising concerns around realism and trust. In this section, we reflect on these results, relate them to prior work on persuasion and interactive systems, and derive implications for the design and evaluation of persuasive technologies and serious games.

\paragraph{\textbf{Delivery format strongly shaped subjective experience even when content was identical.}}
Despite using the same set of arguments and numerical facts across conditions (\S\ref{sec:study-design:topics}), participants' subjective evaluations differed systematically by mode (\S\ref{sec:findings:subjective-outcomes}). The chatbot condition tended to score highest across experience measures (e.g., engagement, enjoyment, trust, and perceived learning), while the essay condition often remained competitive on clarity and ease of following, and the game condition generally trailed. This pattern indicates that differences in subjective experience are not merely a function of informational content, but also of \emph{how} that content is staged, paced, and interacted with. For persuasive-system designers, this underscores that interface and interaction design can be consequential even under strict content control \cite{sundar2008main}.

\paragraph{\textbf{Perceived importance does not immediately lead to behavioral change}}
Among the perceived change and persuasion outcomes (i.e. perceived importance, intention to act, and belief in effectiveness in \S\ref{sec:findings:pre-post-effects}), the clearest mode effect emerged for perceived \emph{importance}. Across the combined-topic models, chat led to significantly stronger increases in importance than essay (and also exceeded game in pairwise models). Although there was a similar weak trend in belief in effectiveness, no comparable effects were observed for intention to act. This pattern suggests that conversational interaction may be especially effective at reframing issue salience -- making a topic feel more personally important -- without immediately altering downstream behavioral intentions. This finding is aligned with classic persuasion accounts in which engagement and perceived relevance can precede, and sometimes decouple from, stable attitude or intention change \citep{PETTY1986123}.

\paragraph{\textbf{Perceived learning and objective retention diverged, especially for the game condition.}}
A key finding is a dissociation between subjective learning judgments and delayed factual recall. Participants in the game condition reported substantially lower self-reported learning than those in essay (\S\ref{sec:findings:subjective-outcomes}), yet achieved higher delayed quiz scores than essay in our objective assessment (\S\ref{sec:findings:objective-knowledge}). One plausible explanation is that interactive formats may support deeper processing and better encoding through active response generation and sustained attention, even when users \emph{feel} that the experience was less informative or less serious \citep{zhonggen2019meta}. In contrast, essays may feel clearer and more ``instructional'' while encouraging comparatively passive consumption. Regardless, this mismatch cautions against relying on self-reported learning as a stand-alone indicator of educational effectiveness in interactive systems. 

\paragraph{\textbf{Realism and trust as a trade-off in narrative game-based persuasion.}}

Participants’ open-ended explanations for changes in perceived importance, intention, and belief largely converged on a common theme: concerns about realism. This theme was mentioned more frequently in the story condition (7 participants) than in the chat (4 participants) and essay (0 participants) conditions. Participants in the story mode specifically noted that the outcomes of each game choice felt ``positive and smooth which is a little unrealistic'', and does not require their ``problem-solving'' mindset. These explanations suggest that some participants interpreted the text-based game as fictional or less grounded in real-world complexity, which in turn may have reduced their trust in the system. This perceived lack of realism not only helps explain the relative lowered persuasiveness scores in the story mode, but also sheds light on the lowered persuasion-related ratings such as convincingness, motivation to act, and influence on thinking, relative to the chat and essay (\S\ref{sec:findings:subjective-outcomes}).
Prior work in narrative persuasion highlights perceived realism and narrative credibility as important moderators of persuasion and counter-arguing \citep{Apple2007,Appel2013,doi:10.1177/0093650212450585}. Our findings align with this literature in suggesting that narrative gamification can support memory and retention, while simultaneously risking a loss of perceived credibility when the framing signals fiction rather than real-world relevance. For designers of serious games, this points to a potential trade-off between the cognitive benefits of interactive narrative and the persuasive benefits of credibility. One implication is that grounding strategies (such as real-world settings, references to credible sources, or reflective prompts linking the experience to players’ own lives) may help preserve trust without sacrificing the benefits of interaction \cite{Angel2015}. 

\paragraph{\textbf{Interaction traces should not be used as proxies for actual learning.}}
Beyond delivery mode effects, our exploratory correlation analyses suggest that interaction structure is more closely related to how participants \emph{experience} an interaction than to what they ultimately retain. In the interaction log analyses (\S\ref{sec:findings:log-analysis}), none of the interaction metrics (e.g., turn counts, word counts, session duration) showed meaningful associations with delayed objective knowledge scores.

In contrast, in the combined chat and game analyses (\S\ref{sec:findings:exploratory:combined}), several interaction properties exhibited consistent directional relationships with \emph{subjective} outcomes. Greater relative user contribution---reflected in longer user messages and higher user--system word ratios---was positively associated with self-reported learning, whereas longer session duration was negatively associated with perceived learning. These patterns mirror structural differences between modes (Table~\ref{tab:interaction-logs-summary}): the game condition involved more extensive turn-taking and more verbose system output, while the chatbot condition elicited more substantial user contributions.

Within-condition analyses (\S\ref{sec:findings:exploratory:separate}) further clarify this distinction. Nominal associations were observed primarily in the chatbot condition, where interaction was open-ended and user-driven, whereas no comparable within-condition associations emerged in the game condition, where interaction structure was more system-controlled. This suggests that interaction traces may be most informative when users have meaningful agency over conversational flow.

Although exploratory and not surviving correction for multiple comparisons, these results offer a clear design signal: behavioral engagement indicators such as verbosity, turn count, or session length should not be assumed to proxy objective learning, even when they shape users’ subjective impressions.

\paragraph{\textbf{Within-chat interaction dynamics offer actionable design signals.}}
While interaction traces should not be treated as indicators of learning outcomes, the within-chat analyses provide useful guidance for the design of interactive persuasive systems. Compared to the game condition---where user input was constrained by narrative structure---the chatbot afforded participants greater control over conversational pacing, resulting in meaningful variability in interaction patterns.

Within the chatbot condition, longer user messages were nominally associated with lower ratings of ease of following, engagement, enjoyment, and motivation to act, whereas a greater number of conversational turns was positively associated with motivation. Together, these patterns suggest that perceived conversational quality depends less on how much is said and more on how interaction is paced and distributed across turns. Sustained back-and-forth may feel motivating, whereas overly long contributions or system-dominated exchanges may increase cognitive load and reduce perceived fluency.

For designers of chat-based persuasive systems, these exploratory findings point to the importance of \emph{interaction balance} rather than maximal engagement---favoring shorter, more frequent exchanges that support incremental user participation. More broadly, they highlight that user agency in conversational pacing may be as important as informational content itself in shaping subjective experience.

\paragraph{\textbf{Design recommendations for persuasive systems and serious games.}}
Lastly, our findings motivate outcome-driven design choices rather than a blanket preference for ``more interactivity''. If the goal is to improve subjective experience and increase perceived importance, conversational delivery appears particularly promising (\S\ref{sec:findings:subjective-outcomes}, \S\ref{sec:findings:pre-post-effects}). If the goal is longer-term factual retention, interactive formats---including narrative game experiences---may provide advantages even when users report learning less (\S\ref{sec:findings:objective-knowledge}). For serious-game developers, two concrete implications follow: (1) \emph{balance user agency and system agency} so that the system does not dominate the interaction through excessive verbosity, and (2) incorporate \emph{credibility and realism cues} (e.g., explicit sourcing, grounded settings, reflection steps) to mitigate the trust penalty that can accompany fictional framing. More broadly, evaluations of persuasive interactive systems should jointly consider self-reported, objective outcomes, and behavioral traces, rather than equating engagement proxies (e.g. turns, verbosity, etc.) with effectiveness.

\section*{Limitations}
\label{sec:limitations}

\paragraph{\textbf{Topic scope.}}
Our study focuses on interactive methods in educational technology for sustainability-related topics. We selected two common topics in this domain -- recycling and public transport. However, there are many other relevant topics, such as renewable energy, waste reduction, and biodiversity. Future studies could examine one or more of these additional topics to provide a more holistic understanding of how interactive methods can be effectively utilized in sustainability education and further benefit society.

\paragraph{\textbf{Participant population and sample size for Measuring Persuasion.}}
We had a total of 45 participants. Because we used a between-subject design, the number of participants within each group was relatively smaller. Participants were recruited through university posters. This may have introduced demographic bias, as the sample consisted primarily of younger and more educated individuals, two demographics which have been shown to hold stronger pre-existing beliefs in climate change \cite{HOEKSTRA2024102851,FURNHAM2022103775}. This pattern was reflected in participants' open-ended responses in the post-study questionnaire. Across all conditions, participants reported prior awareness of the importance and effectiveness of recycling and public transit, as well as pre-existing behavioral intentions. Many indicated that the study simply ``reinforced the idea''. Repeating this study with a different population -- for example, individuals who are more skeptical about and have opposing views on sustainability topics -- may yield different outcomes. Future work looking to expand the population of participants may have to involve other means in the design, such as debunking misinformation about climate change \cite{van2017inoculating}.

\paragraph{\textbf{Knowledge retention and Belief Integration}}
We evaluated knowledge retention after 24 hours, which reflects a short-term retention window. To further validate whether the knowledge is retained in the long term, future work should adopt a longitudinal design and administer follow-up knowledge assessments after a longer delay (e.g., several days or weeks). While remembering information is likely necessary for persuasion to last, it is not sufficient on its own. Remembering more facts does not automatically lead to stronger belief change or behavioral action~\cite{weber2025practice}. In fact, individuals are often unable to escape the pull of their existing attitudes, which guide how new information is processed and evaluated \cite{taber2006motivated}. Future research could therefore examine longer-term retention and test whether retained knowledge translates into durable belief change and behavior.

\section*{Ethical Statement}
\label{sec:ethical-statement}

\noindent \textbf{Access.}
The code base for \ourtool{} is publicly available.

\noindent \textbf{Participant Selection and Compensation.}
Participants were adult volunteers recruited through university advertisements. All participants were fluent in English, ensuring clear communication throughout the study. Participants were compensated \$20 USD for their time, which exceeds local minimum wage standards.

\noindent \textbf{Participant Consent and Data Usage.}
Prior to participation, all participants provided informed consent. The consent form explained the study procedure, data collection, and data usage. Participants were informed that they would be interacting with AI-generated content, advised not to share personal or sensitive information, and assured that all responses would be anonymized and used solely for research purposes.

\noindent \textbf{Ethics Review.}
This study involved a voluntary user study with adult participants. The study protocol was reviewed and approved by the authors’ institutional ethics review board. All participants provided informed consent prior to participation and were informed that they could withdraw from the study at any time without penalty.

\begin{acks}
This work was funded, in part, by the Vector Institute for AI, Canada CIFAR AI Chairs program, Accelerate Foundation Models Research Program
Award from Microsoft, and an NSERC discovery grant. We thank Dongwook Yoon for helpful feedback and discussion.
\end{acks}

\bibliographystyle{ACM-Reference-Format}
\bibliography{references}

@incollection{PETTY1986123,
  author    = {Richard E. Petty and John T. Cacioppo},
  title     = {The Elaboration Likelihood Model of Persuasion},
  booktitle = {Advances in Experimental Social Psychology},
  editor    = {Leonard Berkowitz},
  volume    = {19},
  pages     = {123--205},
  year      = {1986},
  publisher = {Academic Press},
  doi       = {10.1016/S0065-2601(08)60214-2}
}

@article{fernbach2013political,
author = {Philip M. Fernbach and Todd Rogers and Craig R. Fox and Steven A. Sloman},
title ={Political Extremism Is Supported by an Illusion of Understanding},

journal = {Psychological Science},
volume = {24},
number = {6},
pages = {939-946},
year = {2013},
doi = {10.1177/0956797612464058},
    note ={PMID: 23620547},

URL = { 
    
        https://doi.org/10.1177/0956797612464058
    
    

},
eprint = { 
    
        https://doi.org/10.1177/0956797612464058
    
    

}
}

@inproceedings{Adaji2020EvaluatingTS,
  title={Evaluating the Susceptibility of E-commerce Shoppers to Persuasive Strategies. A Game-Based Approach},
  author={Ifeoma Adaji and Nafisul Kiron and Julita Vassileva},
  booktitle={International Conference on Persuasive Technology},
  year={2020},
  url={https://api.semanticscholar.org/CorpusID:215640239}
}

@inproceedings{Adaji2022Review,
author = {Adaji, Ifeoma and Adisa, Mikhail},
title = {A Review of the Use of Persuasive Technologies to Influence Sustainable Behaviour},
year = {2022},
isbn = {9781450392327},
publisher = {Association for Computing Machinery},
address = {New York, NY, USA},
url = {https://doi.org/10.1145/3511047.3537653},
doi = {10.1145/3511047.3537653},
booktitle = {Adjunct Proceedings of the 30th ACM Conference on User Modeling, Adaptation and Personalization},
pages = {317–325},
numpages = {9},
location = {Barcelona, Spain},
series = {UMAP '22 Adjunct}
}

@Article{su141610400,
AUTHOR = {Boncu, Ștefan and Candel, Octav-Sorin and Popa, Nicoleta Laura},
TITLE = {Gameful Green: A Systematic Review on the Use of Serious Computer Games and Gamified Mobile Apps to Foster Pro-Environmental Information, Attitudes and Behaviors},
JOURNAL = {Sustainability},
VOLUME = {14},
YEAR = {2022},
NUMBER = {16},
ARTICLE-NUMBER = {10400},
URL = {https://www.mdpi.com/2071-1050/14/16/10400},
ISSN = {2071-1050},
DOI = {10.3390/su141610400}
}

@inproceedings{folkins2020,
author = {Folkins, Cassandra and Read, Emily and Mundee, Jeff and Birk, Max V. and Bateman, Scott},
title = {A Serious Game for Promoting Positive Attitudes Towards Nursing Homes Among Youth},
year = {2020},
isbn = {9781450380744},
publisher = {Association for Computing Machinery},
address = {New York, NY, USA},
url = {https://doi.org/10.1145/3410404.3414253},
doi = {10.1145/3410404.3414253},
booktitle = {Proceedings of the Annual Symposium on Computer-Human Interaction in Play},
pages = {484–498},
numpages = {15},
location = {Virtual Event, Canada},
series = {CHI PLAY '20}
}

@article{Monroe03062019,
  title={Identifying effective climate change education strategies: A systematic review of the research},
  author={Monroe, Martha C and Plate, Richard R and Oxarart, Annie and Bowers, Alison and Chaves, Willandia A},
  journal={Environmental Education Research},
  volume={25},
  number={6},
  pages={791--812},
  year={2019},
  publisher={Taylor \& Francis}
}

@inproceedings{brynjarsdottir2012sustainably,
  title={Sustainably unpersuaded: how persuasion narrows our vision of sustainability},
  author={Brynjarsdottir, Hronn and H{\aa}kansson, Maria and Pierce, James and Baumer, Eric and DiSalvo, Carl and Sengers, Phoebe},
  booktitle={Proceedings of the sigchi conference on human factors in computing systems},
  pages={947--956},
  year={2012}
}

@Article{app112110089,
AUTHOR = {Murillo-Muñoz, Fernanda and Navarro-Cota, Christian and Juárez-Ramírez, Reyes and Jiménez, Samantha and Nieto Hipólito, Juan Ivan and Molina, Ana I. and Vazquez-Briseno, Mabel},
TITLE = {Characteristics of a Persuasive Educational System: A Systematic Literature Review},
JOURNAL = {Applied Sciences},
VOLUME = {11},
YEAR = {2021},
NUMBER = {21},
ARTICLE-NUMBER = {10089},
URL = {https://www.mdpi.com/2076-3417/11/21/10089},
ISSN = {2076-3417},
DOI = {10.3390/app112110089}
}

@article{kruskal1952use,
  title={Use of ranks in one-criterion variance analysis},
  author={Kruskal, William H and Wallis, W Allen},
  journal={Journal of the American Statistical Association},
  volume={47},
  number={260},
  pages={583--621},
  year={1952},
  publisher={Taylor \& Francis}
}

@article{mann1947test,
  title={On a test of whether one of two random variables is stochastically larger than the other},
  author={Mann, Henry B and Whitney, Donald R},
  journal={The annals of mathematical statistics},
  pages={50--60},
  year={1947},
  publisher={JSTOR}
}

@article{mccullagh1980regression,
  title={Regression models for ordinal data},
  author={McCullagh, Peter},
  journal={Journal of the Royal Statistical Society Series B: Statistical Methodology},
  volume={42},
  number={2},
  pages={109--127},
  year={1980},
  publisher={Oxford University Press}
}

@article{wald1943tests,
  title={Tests of statistical hypotheses concerning several parameters when the number of observations is large},
  author={Wald, Abraham},
  journal={Transactions of the American Mathematical Society},
  volume={54},
  number={3},
  pages={426--482},
  year={1943},
  publisher={JSTOR}
}

@article{cepeda2006distributed,
  title={Distributed practice in verbal recall tasks: A review and quantitative synthesis},
  author={Cepeda, Nicholas J. and Pashler, Harold and Vul, Edward and Wixted, John T. and Rohrer, Doug},
  journal={Psychological Bulletin},
  volume={132},
  number={3},
  pages={354--380},
  year={2006}
}

@article{roediger2006test,
  title={Test-enhanced learning: Taking memory tests improves long-term retention},
  author={Roediger, Henry L. and Karpicke, Jeffrey D.},
  journal={Psychological Science},
  volume={17},
  number={3},
  pages={249--255},
  year={2006}
}

@article{kantak2012learning,
  title={Learning-performance distinction and memory processes for motor skills},
  author={Kantak, Shilpa S. and Winstein, Carole J.},
  journal={Current Opinion in Neurology},
  volume={25},
  number={6},
  pages={590--597},
  year={2012}
}

@article{matthews2016persuasive,
  title={Persuasive technology in mobile applications promoting physical activity: a systematic review},
  author={Matthews, John and Win, Khin Than and Oinas-Kukkonen, Harri and Freeman, Mark},
  journal={Journal of medical systems},
  volume={40},
  number={3},
  pages={72},
  year={2016},
  publisher={Springer}
}

@inproceedings{anderson2013gesture,
  title={Gesture-based interfaces: Learning, performance, and retention},
  author={Anderson, Fraser and Grossman, Tovi and Wigdor, Daniel and Fitzmaurice, George},
  booktitle={Proceedings of the SIGCHI Conference on Human Factors in Computing Systems},
  pages={247--256},
  year={2013}
}

@book{Zar2005,
  author    = {Zar, Jerrold H.},
  title     = {Spearman rank correlation},
  year      = {2005},
  publisher = {Encyclopedia of Biostatistics},
  volume    = {7},
  note      = {Wiley},
  url       = {https://citebay.com/how-to-cite/spearmans-rank-correlation-coefficient/}
}

@article{fernandez2021,
author = {Fern\'{a}ndez Galeote, Daniel and Hamari, Juho},
title = {Game-based Climate Change Engagement: Analyzing the Potential of Entertainment and Serious Games},
year = {2021},
issue_date = {September 2021},
publisher = {Association for Computing Machinery},
address = {New York, NY, USA},
volume = {5},
number = {CHI PLAY},
url = {https://doi.org/10.1145/3474653},
doi = {10.1145/3474653},
journal = {Proc. ACM Hum.-Comput. Interact.},
month = oct,
articleno = {226},
numpages = {21}
}

@article{Nussbaum2015,
  title={Losing the Lake: Simulations to promote gains in student knowledge and interest about climate change.},
  author={Nussbaum, E Michael and Owens, Marissa C and Sinatra, Gale M and Rehmat, Abeera P and Cordova, Jacqueline R and Ahmad, Sajjad and Harris Jr, Fred C and Dascalu, Sergiu M},
  journal={International Journal of Environmental and Science Education},
  volume={10},
  number={6},
  pages={789--811},
  year={2015},
  publisher={ERIC}
}

@inproceedings{Angel2015,
author = {Angel, Jeannette and LaValle, Alicia and Iype, Deepti Mathew and Sheppard, Stephen and Dulic, Aleksandra},
title = {Future delta 2.0 an experiential learning context for a serious game about local climate change},
year = {2015},
isbn = {9781450339278},
publisher = {Association for Computing Machinery},
address = {New York, NY, USA},
url = {https://doi.org/10.1145/2818498.2818512},
doi = {10.1145/2818498.2818512},
booktitle = {SIGGRAPH Asia 2015 Symposium on Education},
articleno = {12},
numpages = {10},
location = {Kobe, Japan},
series = {SA '15}
}

@article{HARKERSCHUCH2020103705,
title = {CO2peration – Structuring a 3D interactive digital game to improve climate literacy in the 12-13-year-old age group},
journal = {Computers \& Education},
volume = {144},
pages = {103705},
year = {2020},
issn = {0360-1315},
doi = {https://doi.org/10.1016/j.compedu.2019.103705},
url = {https://www.sciencedirect.com/science/article/pii/S0360131519302581},
author = {Inez EP Harker-Schuch and Franklin P Mills and Steven J Lade and Rebecca M Colvin}
}

@article{hobert2019say,
  title={Say hello to your new automated tutor--a structured literature review on pedagogical conversational agents},
  author={Hobert, Sebastian and Meyer von Wolff, Raphael},
  year={2019}
}

@article{okonkwo2020python,
  title={Python-bot: A chatbot for teaching python programming.},
  author={Okonkwo, Chinedu Wilfred and Ade-Ibijola, Abejide},
  journal={Engineering Letters},
  volume={29},
  number={1},
  year={2020}
}

@article{Kobi2025,
  title={The levers of political persuasion with conversational artificial intelligence},
  author={Hackenburg, Kobi and Tappin, Ben M and Hewitt, Luke and Saunders, Ed and Black, Sid and Lin, Hause and Fist, Catherine and Margetts, Helen and Rand, David G and Summerfield, Christopher},
  journal={Science},
  volume={390},
  number={6777},
  pages={eaea3884},
  year={2025},
  publisher={American Association for the Advancement of Science}
}

@inproceedings{chisalita2022,
author = {Chisalita, Raluca and Murtinger, Markus and Kriglstein, Simone},
title = {Grow Your Plant: A Plant-Based Game For Creating Awareness About Sustainability Behaviour by Using Renewable Energy},
year = {2022},
isbn = {9781450392112},
publisher = {Association for Computing Machinery},
address = {New York, NY, USA},
url = {https://doi.org/10.1145/3505270.3558344},
doi = {10.1145/3505270.3558344},
booktitle = {Extended Abstracts of the 2022 Annual Symposium on Computer-Human Interaction in Play},
pages = {177–182},
numpages = {6},
location = {Bremen, Germany},
series = {CHI PLAY '22}
}

@article{Fernandez_Galeote_2021,
doi = {10.1088/1748-9326/abec05},
url = {https://doi.org/10.1088/1748-9326/abec05},
year = {2021},
month = {jun},
publisher = {IOP Publishing},
volume = {16},
number = {6},
pages = {063004},
author = {Fernández Galeote, Daniel and Rajanen, Mikko and Rajanen, Dorina and Legaki, Nikoletta-Zampeta and Langley, David J and Hamari, Juho},
title = {Gamification for climate change engagement: review of corpus and future agenda},
journal = {Environmental Research Letters}
}

@article{knol2011enercities,
  title={EnerCities-A serious game to stimulate sustainability and energy conservation: Preliminary results},
  author={Knol, Erik and De Vries, Peter W},
  journal={eLearning Papers},
  number={25},
  year={2011}
}

@inproceedings{Newsome2020,
author = {Newsome, Isabel},
title = {An Educational Game Bringing Awareness to Declining Insect Populations},
year = {2020},
isbn = {9781450375870},
publisher = {Association for Computing Machinery},
address = {New York, NY, USA},
url = {https://doi.org/10.1145/3383668.3419912},
doi = {10.1145/3383668.3419912},
booktitle = {Extended Abstracts of the 2020 Annual Symposium on Computer-Human Interaction in Play},
pages = {326–329},
numpages = {4},
location = {Virtual Event, Canada},
series = {CHI PLAY '20}
}

@inproceedings{Troiano2020,
author = {Troiano, Giovanni Maria and Schouten, Dylan and Cassidy, Michael and Tucker-Raymond, Eli and Puttick, Gillian and Harteveld, Casper},
title = {Ice Paddles, CO2 Invaders, and Exploding Planets: How Young Students Transform Climate Science Into Serious Games},
year = {2020},
isbn = {9781450380744},
publisher = {Association for Computing Machinery},
address = {New York, NY, USA},
url = {https://doi.org/10.1145/3410404.3414256},
doi = {10.1145/3410404.3414256},
booktitle = {Proceedings of the Annual Symposium on Computer-Human Interaction in Play},
pages = {534–548},
numpages = {15},
location = {Virtual Event, Canada},
series = {CHI PLAY '20}
}

@article{Persky2020PerceptionOL,
  title={Perception of Learning Versus Performance as Outcome Measures of Educational Research},
  author={Adam M. Persky and Edward Lee and Lauren S. Schlesselman},
  journal={American Journal of Pharmaceutical Education},
  year={2020},
  volume={84},
  url={https://api.semanticscholar.org/CorpusID:216154214}
}

@article{AlAperaparSingh2023THEDO,
  title={The dichotomization of objective and subjective outcome measures of academic performance in an online learning environment},
  author={Aperapar Singh, Parmjit Singh and Anthonysamy, Lilian},
  journal={Malaysian Journal of Learning and Instruction (MJLI)},
  volume={20},
  number={1},
  pages={63--92},
  year={2023},
  publisher={Universiti Utara Malaysia Press}
}

@inproceedings{furumai-etal-2024-zero,
    title = "Zero-shot Persuasive Chatbots with {LLM}-Generated Strategies and Information Retrieval",
    author = "Furumai, Kazuaki  and
      Legaspi, Roberto  and
      Romero, Julio Cesar Vizcarra  and
      Yamazaki, Yudai  and
      Nishimura, Yasutaka  and
      Semnani, Sina  and
      Ikeda, Kazushi  and
      Shi, Weiyan  and
      Lam, Monica",
    editor = "Al-Onaizan, Yaser  and
      Bansal, Mohit  and
      Chen, Yun-Nung",
    booktitle = "Findings of the Association for Computational Linguistics: EMNLP 2024",
    month = nov,
    year = "2024",
    address = "Miami, Florida, USA",
    publisher = "Association for Computational Linguistics",
    url = "https://aclanthology.org/2024.findings-emnlp.656/",
    doi = "10.18653/v1/2024.findings-emnlp.656",
    pages = "11224--11249"
}

@book{mercier2019enigma,
  title={The Enigma of Reason},
  author={Mercier, H. and Sperber, D.},
  isbn={9780674237827},
  lccn={2016050637},
  url={https://books.google.ca/books?id=6JeDvQEACAAJ},
  year={2019},
  publisher={Harvard University Press}
}

@book{mercier2020notborn,
  title     = {Not Born Yesterday: The Science of Who We Trust and What We Believe},
  author    = {Mercier, Hugo},
  year      = {2020},
  publisher = {Princeton University Press},
  address   = {Princeton, NJ}
}

@article{luciano2024hypersuasion,
  title={Hypersuasion--On AI’s persuasive power and how to deal with it},
  author={Luciano, Floridi},
  journal={Philosophy \& Technology},
  volume={37},
  number={2},
  pages={64},
  year={2024},
  publisher={Springer}
}

@online{durmus2024persuasion,
author = {Esin Durmus and Liane Lovitt and Alex Tamkin and Stuart Ritchie and Jack Clark and Deep Ganguli},
title = {Measuring the Persuasiveness of Language Models},
date = {2024-04-09},
year = {2024},
url = {https://www.anthropic.com/news/measuring-model-persuasiveness},
}

@article{salvi2025conversational,
  title={On the conversational persuasiveness of GPT-4},
  author={Salvi, Francesco and Horta Ribeiro, Manoel and Gallotti, Riccardo and West, Robert},
  journal={Nature Human Behaviour},
  pages={1--9},
  year={2025},
  publisher={Nature Publishing Group UK London}
}

@article{bai2025llm,
  title={LLM-generated messages can persuade humans on policy issues},
  author={Bai, Hui and Voelkel, Jan G and Muldowney, Shane and Eichstaedt, Johannes C and Willer, Robb},
  journal={Nature Communications},
  volume={16},
  number={1},
  pages={6037},
  year={2025},
  publisher={Nature Publishing Group UK London}
}

@article{matz2024potential,
  title={The potential of generative AI for personalized persuasion at scale},
  author={Matz, Sandra C and Teeny, Jacob D and Vaid, Sumer S and Peters, Heinrich and Harari, Gabriella M and Cerf, Moran},
  journal={Scientific Reports},
  volume={14},
  number={1},
  pages={4692},
  year={2024},
  publisher={Nature Publishing Group UK London}
}

@article{pnasnexus,
  title={The persuasive effects of political microtargeting in the age of generative artificial intelligence},
  author={Simchon, Almog and Edwards, Matthew and Lewandowsky, Stephan},
  journal={PNAS nexus},
  volume={3},
  number={2},
  pages={pgae035},
  year={2024},
  publisher={Oxford University Press US}
}

@article{rogiers2024persuasion,
  title={Persuasion with large language models: a survey},
  author={Rogiers, Alexander and Noels, Sander and Buyl, Maarten and De Bie, Tijl},
  journal={arXiv preprint arXiv:2411.06837},
  year={2024}
}

@article{schoenegger2025large,
  title={Large Language Models Are More Persuasive Than Incentivized Human Persuaders},
  author={Schoenegger, Philipp and Salvi, Francesco and Liu, Jiacheng and Nan, Xiaoli and Debnath, Ramit and Fasolo, Barbara and Leivada, Evelina and Recchia, Gabriel and G{\"u}nther, Fritz and Zarifhonarvar, Ali and others},
  journal={arXiv preprint arXiv:2505.09662},
  year={2025}
}

@article{nam2025llms,
  title={LLMs as Scalable Tools for Interactive Consumer Behavior Experiments: Comparing Persuasion Strategy Effectiveness},
  author={Nam, Jimin and Orchinik, Reed and Rand, David G},
  year={2025},
  publisher={OSF}
}

@article{altay2023information,
  title={Information delivered by a chatbot has a positive impact on COVID-19 vaccines attitudes and intentions.},
  author={Altay, Sacha and Hacquin, Anne-Sophie and Chevallier, Coralie and Mercier, Hugo},
  journal={Journal of Experimental Psychology: Applied},
  volume={29},
  number={1},
  pages={52},
  year={2023},
  publisher={American Psychological Association}
}

@article{salvi2024conversational,
  title={On the conversational persuasiveness of large language models: A randomized controlled trial},
  author={Salvi, Francesco and Ribeiro, Manoel Horta and Gallotti, Riccardo and West, Robert},
  journal={arXiv preprint arXiv:2403.14380},
  year={2024}
}

@article{karinshak2023working,
  title={Working with AI to persuade: Examining a large language model's ability to generate pro-vaccination messages},
  author={Karinshak, Elise and Liu, Sunny Xun and Park, Joon Sung and Hancock, Jeffrey T},
  journal={Proceedings of the ACM on Human-Computer Interaction},
  volume={7},
  number={CSCW1},
  pages={1--29},
  year={2023},
  publisher={ACM New York, NY, USA}
}

@article{goldstein2024persuasive,
  title={How persuasive is AI-generated propaganda?},
  author={Goldstein, Josh A and Chao, Jason and Grossman, Shelby and Stamos, Alex and Tomz, Michael},
  journal={PNAS nexus},
  volume={3},
  number={2},
  pages={pgae034},
  year={2024},
  publisher={Oxford University Press US}
}

@article{nazry2017mood,
  title={Mood and learning in navigation-based serious games},
  author={Nazry, N Nazrina M and Romano, Daniela M},
  journal={Computers in Human Behavior},
  volume={73},
  pages={596--604},
  year={2017},
  publisher={Elsevier}
}

@book{de2017persuasive,
  title={Persuasive Technology: Development and Implementation of Personalized Technologies to Change Attitudes and Behaviors: 12th International Conference, PERSUASIVE 2017, Amsterdam, The Netherlands, April 4--6, 2017, Proceedings},
  author={De Vries, Peter W and Oinas-Kukkonen, Harri and Siemons, Liseth and Beerlage-de Jong, Nienke and van Gemert-Pijnen, Lisette},
  volume={10171},
  year={2017},
  publisher={Springer}
}

@inproceedings{chockkalingam-etal-2025-go,
    title = "Should {I} go vegan: Evaluating the Persuasiveness of {LLM}s in Persona-Grounded Dialogues",
    author = "Chockkalingam, Shruthi  and
      Hossein Alavi, Seyed  and
      T. Ng, Raymond  and
      Shwartz, Vered",
    editor = "Hale, James  and
      Kwon, Brian Deuksin  and
      Dutt, Ritam",
    booktitle = "Proceedings of the Third Workshop on Social Influence in Conversations (SICon 2025)",
    month = jul,
    year = "2025",
    address = "Vienna, Austria",
    publisher = "Association for Computational Linguistics",
    url = "https://aclanthology.org/2025.sicon-1.4/",
    doi = "10.18653/v1/2025.sicon-1.4",
    pages = "65--72",
    ISBN = "979-8-89176-266-4"
}

@inproceedings{elmgadmi2025large,
  title={Large Language Models and Non-Player Characters in Gaming: A Bibliometric Overview},
  author={Elmgadmi, Hoda and Nafil, Khalid},
  booktitle={2025 International Conference on Intelligent Systems: Theories and Applications (SITA)},
  pages={1--8},
  year={2025},
  organization={IEEE}
}

@article{shields2025could,
  title={Could vs Should: Exploring Prompting Strategies and Writer Perspectives Towards LLM Assistance in Storylet Authoring},
  author={Shields, Samuel and Lafosse, Celine and Johnson-Bey, Shi and Hwang, Daeun and Wardrip-Fruin, Noah and Melcer, Edward F},
  journal={IEEE Transactions on Games},
  year={2025},
  publisher={IEEE}
}

@article{alavi2024mcpdial,
  title={Mcpdial: A minecraft persona-driven dialogue dataset},
  author={Alavi, Seyed Hossein and Rao, Sudha and Adhikari, Ashutosh and DesGarennes, Gabriel A and Malhotra, Akanksha and Brockett, Chris and Adada, Mahmoud and Ng, Raymond T and Shwartz, Vered and Dolan, Bill},
  journal={arXiv preprint arXiv:2410.21627},
  year={2024}
}

@article{alavi2024game,
  title={Game Plot Design with an LLM-powered Assistant: An Empirical Study with Game Designers},
  author={Alavi, Seyed Hossein and Xu, Weijia and Jojic, Nebojsa and Kennett, Daniel and Ng, Raymond T and Rao, Sudha and Zhang, Haiyan and Dolan, Bill and Shwartz, Vered},
  journal={arXiv preprint arXiv:2411.02714},
  year={2024}
}

@article{yang2023does,
  title={How does interactive virtual reality enhance learning outcomes via emotional experiences? A structural equation modeling approach},
  author={Yang, Hairu and Cai, Minghan and Diao, Yongfeng and Liu, Rui and Liu, Ling and Xiang, Qianchen},
  journal={Frontiers in Psychology},
  volume={13},
  pages={1081372},
  year={2023},
  publisher={Frontiers Media SA}
}

@article{zhonggen2019meta,
  title={A meta-analysis of use of serious games in education over a decade},
  author={Zhonggen, Yu},
  journal={International Journal of Computer Games Technology},
  volume={2019},
  number={1},
  pages={4797032},
  year={2019},
  publisher={Wiley Online Library}
}

@article{doi:10.1177/0093650212450585,
  title={Perceived realism: Dimensions and roles in narrative persuasion},
  author={Cho, Hyunyi and Shen, Lijiang and Wilson, Kari},
  journal={Communication research},
  volume={41},
  number={6},
  pages={828--851},
  year={2014},
  publisher={Sage Publications Sage CA: Los Angeles, CA}
}

@article{Appel2013,
  title={The persuasive influence of a fictional character's trustworthiness},
  author={Appel, Markus and Mara, Martina},
  journal={Journal of Communication},
  volume={63},
  number={5},
  pages={912--932},
  year={2013},
  publisher={Oxford University Press}
}

@article{Apple2007,
author = {Appel, Markus and Richter, Tobias},
year = {2007},
month = {12},
pages = {113-134},
title = {Persuasive Effects of Fictional Narratives Increase Over Time},
volume = {10},
journal = {Media Psychology},
doi = {10.1080/15213260701301194}
}

@article{benjamini1995controlling,
  title={Controlling the False Discovery Rate: A Practical and Powerful Approach to Multiple Testing},
  author={Benjamini, Yoav and Hochberg, Yosef},
  journal={Journal of the Royal Statistical Society: Series B (Methodological)},
  volume={57},
  number={1},
  pages={289--300},
  year={1995},
  publisher={Wiley}
}

@article{HOEKSTRA2024102851,
title = {The educational divide in climate change attitudes: Understanding the role of scientific knowledge and subjective social status},
journal = {Global Environmental Change},
volume = {86},
pages = {102851},
year = {2024},
issn = {0959-3780},
doi = {https://doi.org/10.1016/j.gloenvcha.2024.102851},
url = {https://www.sciencedirect.com/science/article/pii/S0959378024000554},
author = {Anne G. Hoekstra and Kjell Noordzij and Willem {de Koster} and Jeroen {van der Waal}}
}

@article{FURNHAM2022103775,
title = {Correlates of belief in climate change: Demographics, ideology and belief systems},
journal = {Acta Psychologica},
volume = {230},
pages = {103775},
year = {2022},
issn = {0001-6918},
doi = {https://doi.org/10.1016/j.actpsy.2022.103775},
url = {https://www.sciencedirect.com/science/article/pii/S0001691822002906},
author = {Adrian Furnham and Charlotte Robinson}
}

@article{van2017inoculating,
  title={Inoculating the public against misinformation about climate change},
  author={Van der Linden, Sander and Leiserowitz, Anthony and Rosenthal, Seth and Maibach, Edward},
  journal={Global challenges},
  volume={1},
  number={2},
  pages={1600008},
  year={2017},
  publisher={Wiley Online Library}
}

@article{weber2025practice,
  title={Practice testing enhances learning but not attitude change from persuasive texts},
  author={Weber, Elena M Galeano and Lehnen, Lisa and Lombardi, Doug and Brod, Garvin},
  journal={Scientific Reports},
  volume={15},
  number={1},
  pages={32935},
  year={2025},
  publisher={Nature Publishing Group UK London}
}

@article{taber2006motivated,
  title={Motivated skepticism in the evaluation of political beliefs},
  author={Taber, Charles S and Lodge, Milton},
  journal={American journal of political science},
  volume={50},
  number={3},
  pages={755--769},
  year={2006},
  publisher={Wiley Online Library}
}

@incollection{sundar2008main,
  title={The MAIN Model: A Heuristic Approach to Understanding Technology Effects on Credibility},
  author={Sundar, S. Shyam},
  booktitle={Digital Media and Learning},
  editor={Metzger, Miriam J. and Flanagin, Andrew J.},
  year={2008},
  publisher={MacArthur Foundation/MIT Press},
  pages={73--100},
  url={https://www.issuelab.org/resource/the-main-model-a-heuristic-approach-to-understanding-technology-effects-on-credibility.html}
}

@inproceedings{10.1145/2207676.2208394,
author = {Thieme, Anja and Comber, Rob and Miebach, Julia and Weeden, Jack and Kraemer, Nicole and Lawson, Shaun and Olivier, Patrick},
title = {"We've bin watching you": designing for reflection and social persuasion to promote sustainable lifestyles},
year = {2012},
isbn = {9781450310154},
publisher = {Association for Computing Machinery},
address = {New York, NY, USA},
url = {https://doi.org/10.1145/2207676.2208394},
doi = {10.1145/2207676.2208394},
booktitle = {Proceedings of the SIGCHI Conference on Human Factors in Computing Systems},
pages = {2337–2346},
numpages = {10},
location = {Austin, Texas, USA},
series = {CHI '12}
}

@String{Computing = "Computing" }

@String{Computer = "{IEEE} Computer" }

@String{Academic = "Academic Press" }

@String{Springer = "Springer-Verlag" }

\appendix
\section{User Study Questionnaires}
\label{appendix:questionnaires}

\subsection{Pre-Study Questionnaire}
\label{appendix:questionnaires:pre-study}

Table~\ref{tab:pre-study-questionnaire} summarizes the pre-study questionnaire administered prior to the experiment. The questionnaire collected demographic and contextual information, as well as participants’ self-reported attitudes, behaviors, and confidence related to the assigned topic (recycling or public transit).


\begin{table*}[t]
\centering
\scriptsize
\setlength{\tabcolsep}{4pt}
\renewcommand{\arraystretch}{1.08}
\begin{tabularx}{\textwidth}{p{3.6cm} X p{3.4cm}}
\toprule
\textbf{Category} & \textbf{Question} & \textbf{Response format}  \\
\midrule

\textbf{Demographics} &
What is your age? &
Categorical (18--24, 25--34, 35--44, 45--54, 55--64, 65+) \\

\midrule

\textbf{Topic assignment} &
What is the topic that you were assigned? &
Multiple choice\\

\midrule

\textbf{Context (Recycling)} &
Do you have convenient access to recycling services at home or in your building? &
Yes / No\\

\midrule

\textbf{Importance (Recycling)} &
How important is recycling to you personally? &
Likert (1--5)  \\

\textbf{Behavioral intention (Recycling)} &
How often do you recycle materials such as paper, plastic, glass, or metal? &
Categorical; recoded to ordinal 1--5 (Never $\rightarrow$ Always) \\

\textbf{Epistemic conficence (Recycling)} &
How confident are you in your knowledge about how recycling works and its impact on reducing waste and protecting the environment? &
Likert (1--5) \\

\midrule

\textbf{Context (Transit)} &
Do you have regular access to a personal car? &
Yes / No  \\

\midrule

\textbf{Importance (Transit)} &
How important is choosing public transit instead of personal cars to you personally? &
Likert (1--5) \\

\textbf{Behaviroal intention (Transit)} &
On a typical week, how many days do you commute outside the home? &
Ordinal Categorical; recoded to ordinal 1--4 (0, 1--2, 3--4, 5+)  \\

\textbf{Epistemic confidence(Transit)} &
How confident are you in your knowledge about the environmental and social benefits of using public transit instead of personal cars? &
Likert (1--5) \\

\bottomrule
\end{tabularx}
\caption{Pre-study questionnaire items collected for descriptive and contextual purposes.}
\label{tab:pre-study-questionnaire}
\end{table*}

\subsection{Post-Study Questionnaire}
\label{appendix:questionnaires:post-study}

Table~\ref{tab:post-study-questionnaire} presents the post-study questionnaire administered immediately after the experimental session. The questionnaire captured participants' subjective experience (e.g., easy to follow, engagement, self-reported learning, enjoyment, trust, influence), and topic-specific perceived changes in importance, intention to act, and belief in effectiveness. Subjective experience items were measured using 5-point Likert scales. Although ``convincingness of arguments'' was collected as a topic-specific item using a slightly different Likert wording (Not at all–Very), we report it alongside other subjective outcomes because it captures participants’ experiential judgment of the persuasive content rather than objective belief change.

Perceived-change items used ordered categorical responses (\textit{Less / Same / More / Not sure}) and were recoded to ordinal values (-1, 0, +1) for analysis, with \textit{Not sure} responses treated as missing. Open-ended prompts were included to collect brief qualitative reflections. 


\begin{table*}[t]
\centering
\scriptsize
\setlength{\tabcolsep}{4pt}
\renewcommand{\arraystretch}{1.05}
\begin{tabularx}{\textwidth}{p{2.6cm} X p{2.6cm} p{2cm}}
\toprule
\textbf{Category} & \textbf{Question} & \textbf{Response format} & \textbf{Applies to} \\
\midrule

\textbf{Subjective experience} & & &\\

\textit{Ease of following} & The information was presented in a way that was easy to follow. &
Likert (1=SD, 5=SA) & All topics \\

\textit{Engagement} & I felt engaged while taking part in this experience. &
Likert (1=SD, 5=SA) & All topics \\

\textit{Self-reported learnin}g & This experience increased my understanding/knowledge of the topic. &
Likert (1=SD, 5=SA) & All topics \\

\textit{Enjoyment} & I enjoyed the way this experience delivered the information. &
Likert (1=SD, 5=SA) & All topics \\

\textit{Trust} & I trusted the information provided in this experience. &
Likert (1=SD, 5=SA) & All topics \\

\textit{Would Recommend} & I would recommend this kind of experience to other people. &
Likert (1=SD, 5=SA) & All topics \\

\textit{Re-encounter in future} & I would like to encounter this type of experience again in the future. &
Likert (1=SD, 5=SA) & All topics \\

\textit{Increased Motivation to act}& This experience increased my motivation to act more sustainably. &
Likert (1=SD, 5=SA) & All topics \\

\textit{Influenced thinking}& Overall, this experience influenced the way I think about this topic. &
Likert (1=SD, 5=SA) & All topics \\

\midrule

\textbf{Topic assignment} &
What is the topic that you were assigned? &
Multiple choice & All topics \\

\midrule

\textbf{Perceived change (Recycling)} & & & \\
Importance change & Compared to before this session, how has the personal importance of recycling changed for you? &
Less / Same / More / Not sure & Recycling only \\

Behavioral intention change & Compared to before this session, how has your intention to recycle whenever possible changed? &
Less / Same / More / Not sure & Recycling only \\

Belief in effectiveness change & Compared to before this session, how has your belief in the effectiveness of recycling changed? &
Less / Same / More / Not sure & Recycling only \\

\midrule

\textbf{Convincingness (Recycling)} &
How convincing did you find the arguments about recycling? &
Likert (1=Not at all, 5=Very) & Recycling only \\

\midrule

\textbf{Perceived change (Transit)} & & & \\

Importance change & Compared to before this session, how has the personal importance of choosing public transit changed? &
Less / Same / More / Not sure & Transit only \\

Behavioral intention change& Compared to before this session, how has your intention to use public transit instead of a personal car changed? &
Less / Same / More / Not sure & Transit only \\
Belief in effectiveness change & Compared to before this session, how has your belief in the effectiveness of public transit changed? &
Less / Same / More / Not sure & Transit only \\

\midrule

\textbf{Convincingness (Transit)} &
How convincing did you find the arguments about using public transit instead of personal cars? &
Likert (1=Not at all, 5=Very) & Transit only \\

\midrule

\textbf{Open-ended reflections} &
Why? Please provide brief explanation (2--3 sentences). &
Free text & After each perceived change/convincing item \\

& Please describe how (if at all) this experience influenced your thoughts, feelings, or actions related to the topic (3--4 sentences). &
Free text & All topics \\

\bottomrule
\end{tabularx}
\caption{Post-study questionnaire used to acquire subjective experience of participants.}

\label{tab:post-study-questionnaire}
\end{table*}

\subsection{Objective Knowledge Quiz}
\label{appendix:questionnaires:knowledge-quiz}

Tables ~\ref{tab:knowledge-quiz-recycling} and ~\ref{tab:knowledge-quiz-transit} list the multiple-choice questions used to assess objective knowledge retention. For each topic, participants answered five questions whose answers were explicitly covered during the experiment session, along with two control questions referencing information not presented. Control questions were included to discourage guessing or external lookup; selecting “I have not seen this information before” was treated as the correct response. Knowledge scores were computed as the number of correct responses to the five content-covered questions only. Confidence ratings collected after each question (via ``How confident are you in your answer to the previous question?'' question) were not analyzed in this work.


\begin{table*}[t]
\centering
\scriptsize
\setlength{\tabcolsep}{4pt}
\renewcommand{\arraystretch}{1.05}
\begin{tabularx}{\textwidth}{p{5cm} X p{2.5cm} p{2cm}}
\toprule
\textbf{Question} & \textbf{Answer options} & \textbf{Correct answer} & \textbf{Type} \\
\midrule

Recycling one ton of paper saves approximately how many trees? &
10; 13; 17; 20; I have not seen this information before &
17 &
Content \\

Recycling aluminum saves up to what percentage of energy compared to producing new aluminum? &
80\%; 85\%; 90\%; 95\%; I have not seen this information before &
95\% &
Content \\

Recycling just one aluminum can can power a TV for approximately how long? &
30 minutes; 1 hour; 3 hours; 5 hours; I have not seen this information before &
I have not seen this information before &
Control \\

How many more jobs does the recycling industry generate compared to landfill management? &
10$\times$ more; 5$\times$ more; 2$\times$ more; 7$\times$ more; I have not seen this information before &
10$\times$ more &
Content \\

Reducing landfill waste can lower disposal costs by up to what percentage? &
10\%; 20\%; 30\%; 40\%; I have not seen this information before &
I have not seen this information before &
Control \\

Unrecycled plastic breaks into microplastics that harm how many marine species? &
Over 70; Over 300; Over 500; Over 700; I have not seen this information before &
Over 700 &
Content \\

Companies with recycling programs see up to what percentage increase in employee engagement and retention? &
10\%; 15\%; 20\%; 25\%; I have not seen this information before &
20\%&
Content \\
\midrule
How confident are you in your answer to the previous question? & Likert 1-5 (Not confident at all $\rightarrow$ Extremely Confident) & N/A & N/A\\
\bottomrule
\end{tabularx}
\caption{Objective knowledge quiz items for the recycling topic. Each participant answered five content-covered questions and two control questions. Control questions referenced information not presented during the experiment; selecting ``I have not seen this information before'' was considered the correct response. After each question (content or control), particpants were asked a confidence question ``How confident are you in your answer to the previous question''.}
\label{tab:knowledge-quiz-recycling}
\end{table*}


\begin{table*}[t]
\centering
\scriptsize
\setlength{\tabcolsep}{4pt}
\renewcommand{\arraystretch}{1.05}
\begin{tabularx}{\textwidth}{p{5cm} X p{2.5cm} p{2cm}}
\toprule
\textbf{Question} & \textbf{Answer options} & \textbf{Correct answer} & \textbf{Type} \\
\midrule

How much can switching from personal cars to public transit reduce an individual’s transportation emissions? &
20\%; 35\%; 50\%; 75\%; I have not seen this information before &
50\% &
Content \\

A 10\% shift from personal cars to public transit can cut commute times by up to: &
15\%; 25\%; 40\%; 60\%; I have not seen this information before &
40\% &
Content \\

How much do businesses typically spend per employee on parking alone each year? &
\$500-\$1{,}000; \$1{,}000-\$1{,}500; \$1{,}500-\$2{,}000; \$2{,}500-\$3{,}000; I have not seen this information before &
I have not seen this information before &
Control \\

How many people can a single lane accommodate when used by buses or trains instead of cars? &
2{,}500; 10{,}000; 15{,}000; 20{,}000; I have not seen this information before &
I have not seen this information before &
Control \\

On average, how much money can a commuter save per year by not driving and using public transit instead? &
\$3{,}000; \$5{,}000; \$7{,}500; \$10{,}000+; I have not seen this information before &
\$10{,}000+ &
Content \\

How much safer is traveling by public transit compared to driving a personal car? &
5$\times$ safer; 10$\times$ safer; 20$\times$ safer; 40$\times$ safer; I have not seen this information before &
10$\times$ safer &
Content \\

Every \$1 invested in public transit generates approximately how much in community benefits? &
\$4; \$5; \$6; \$7; I have not seen this information before &
\$4 &
Content \\

\midrule
How confident are you in your answer to the previous question? &
Likert 1--5 (Not confident at all $\rightarrow$ Extremely confident) &
N/A &
N/A \\
\bottomrule
\end{tabularx}

\caption{Objective knowledge quiz items for the public transit topic. Each participant answered five content-covered questions and two control questions. Control questions referenced information not presented during the experiment; selecting ``I have not seen this information before'' was considered the correct response. After each question (content or control), participants were asked a confidence question.}
\label{tab:knowledge-quiz-transit}
\end{table*}

\section{Key Arguments}
\label{appendix:key-arguments}
For each topic (recycling and public transit), we compiled five persuasive arguments, each paired with a concrete numerical fact.
The complete argument sets are listed in Tables~\ref{tab:arguments-recycling-transit}.
All prompts referencing \texttt{<key\_arguments>} loaded these files verbatim. 

\begin{table}[t]
    \centering
    \scriptsize
    \setlength{\tabcolsep}{3.2pt}
    \tt
    \begin{tabular}{|p{0.9\textwidth}|}
    \hline
    
\rowcolor{gray!10}\multicolumn{1}{|c|}{\textbf{\textsf{Topic A: Recycling}}}\\ \hline

Here are five compelling arguments to persuade individuals and businesses to adopt better recycling habits:

\vspace{0.4em}

\textcolor{blue}{\textbf{1. Environmental Preservation: Protecting Natural Resources and Ecosystems}}

\vspace{0.2em}

\textcolor{gray}{\textbf{Premise:}}

Recycling reduces raw material extraction, preserving forests, water, and ecosystems.

\textcolor{cadmiumgreen}{\textbf{Fact:} Recycling one ton of paper saves 17 trees.}

\vspace{0.6em}

\textcolor{blue}{\textbf{2. Energy Conservation and Lower Carbon Emissions}}

\vspace{0.2em}

\textcolor{gray}{\textbf{Premise:}}

Recycling uses much less energy than making new materials, cutting emissions.

\textcolor{cadmiumgreen}{\textbf{Fact:} Recycling aluminum saves up to 95\% of the energy compared to new production.}

\vspace{0.6em}

\textcolor{blue}{\textbf{3. Economic Benefits: Cost Savings and Job Creation}}

\vspace{0.2em}

\textcolor{gray}{\textbf{Premise:}}

Recycling creates more jobs and saves businesses money on waste.

\textcolor{cadmiumgreen}{\textbf{Fact:} Recycling creates 10 jobs for every landfill job.}

\vspace{0.6em}

\textcolor{blue}{\textbf{4. Reducing Pollution: Cleaner Air, Water, and Soil}}

\vspace{0.2em}

\textcolor{gray}{\textbf{Premise:}}

Recycling reduces pollution from landfill waste and incineration.

\textcolor{cadmiumgreen}{\textbf{Fact:} Unrecycled plastic breaks into microplastics that harm over 700 marine species and enter the food chain.}

\vspace{0.6em}

\textcolor{blue}{\textbf{5. Setting a Positive Example and Meeting Social Responsibility Goals}}

\vspace{0.2em}

\textcolor{gray}{\textbf{Premise:}}

Recycling shows commitment and inspires sustainable behavior.

\textcolor{cadmiumgreen}{\textbf{Fact:} Companies with recycling programs have up to 20\% higher employee engagement and retention.}

\vspace{0.8em}

By highlighting these arguments, you can effectively appeal to the environmental, economic, and social benefits of better recycling habits, making the case for why change is both necessary and rewarding.\\
    \hline
    \rowcolor{gray!10}\multicolumn{1}{|c|}{\textbf{\textsf{Topic B: Public Transit}}}\\ \hline

    Here are five key arguments for convincing individuals and businesses to use public transit instead of personal cars:

\vspace{0.4em}

\textcolor{blue}{\textbf{1. Environmental Benefits: Reduced Carbon Footprint}}

\vspace{0.2em}

\textcolor{gray}{\textbf{Premise:}}

Public transit cuts emissions and helps clean the air.

\textcolor{cadmiumgreen}{\textbf{Fact:} Switching to transit can cut your transportation emissions by 50\%.}

\vspace{0.6em}

\textcolor{blue}{\textbf{2. Cost Savings for Individuals and Businesses}}

\vspace{0.2em}

\textcolor{gray}{\textbf{Premise:}}

Public transit saves money for commuters and employers.

\textcolor{cadmiumgreen}{\textbf{Fact:} The average commuter saves \$10{,}000+ per year by not driving.}

\vspace{0.6em}

\textcolor{blue}{\textbf{3. Reduced Traffic Congestion}}

\vspace{0.2em}

\textcolor{gray}{\textbf{Premise:}}

Transit reduces traffic and speeds up commutes.

\textcolor{cadmiumgreen}{\textbf{Fact:} A 10\% shift to transit can cut commute times by up to 40\%.}

\vspace{0.6em}

\textcolor{blue}{\textbf{4. Health and Safety Improvements}}

\vspace{0.2em}

\textcolor{gray}{\textbf{Premise:}}

Transit is safer and helps reduce pollution-related illness.

\textcolor{cadmiumgreen}{\textbf{Fact:} Transit is 10$\times$ safer than driving.}

\vspace{0.6em}

\textcolor{blue}{\textbf{5. Economic and Social Benefits for Communities}}

\vspace{0.2em}

\textcolor{gray}{\textbf{Premise:}}

Transit boosts local economies and access for all.

\textcolor{cadmiumgreen}{\textbf{Fact:}  Every \$1 spent on transit returns \$4 in community benefits.}

\vspace{0.8em}

By emphasizing these arguments, individuals and businesses can see the tangible benefits of public transit, not just for themselves, but for their communities and the environment as a whole.\\

\hline
    \end{tabular}
    \caption{Fixed persuasive arguments and supporting facts for the recycling (top section) and transit (bottom section). Each argument consists of a high-level premise paired with a numerical fact. This content was made by the authors of this paper and used verbatim across all experiment
conditions (i.e. essay, chat, and game).}
    \label{tab:arguments-recycling-transit}
\end{table}

\section{Essay Generation}
\label{appendix:essay-generation-prompt}
Table \ref{tab:essay-generation-prompt} shows the template prompt used for generating essays in our experiment. Essay generation was performed using GPT-4.1 with a maximum completion length of 33k tokens and default decoding parameters. We generated 20 essays per topic in advance and sampled from this fixed set during the experiment to reduce interaction delays and avoid unforeseen technical issues during data collection.

\begin{table}[h]
    \centering
    \scriptsize
    \setlength{\tabcolsep}{3.2pt}
    \tt
    \begin{tabular}{|p{0.9\textwidth}|}
    \hline
        \textcolor{red}{System Message:} You are an expert essay writer. For each key argument write a paragraph and focus on the provided fact. The essay must follow the following formatting:\\
    <article>\\
    <h3><strong>Essay Title</strong></h3>\\
    <p>paragraph 1</p>\\
    <p>paragraph 2</p>\\
    ...\\
    <p> Paragraph n</p>\\
    </article>\\
        
        \textcolor{red}{User Message:} Write a persuasive and informative essay based on the following key arguments:\\
        \textcolor{cadmiumgreen}{<key arguments>}\\
    \hline
    \end{tabular}
    \caption{Template prompt used for generating essays. We generated 20 prompts for each topic (i.e. recycling and transit), and sampled from them during our user study. See Appendix \ref{appendix:key-arguments} for more details on key arguments. }
    \label{tab:essay-generation-prompt}
\end{table}

\section{Chatbot Implementation}
\label{appendix:chatbot-implementation}

The chatbot condition was implemented using a large language model with structured system and user prompts. Table~\ref{tab:chat-interaction-prompt} illustrates the construction of the system and user messages, along with the initial static prompts used for the recycling and public transit topics. Throughout the experiment we used GPT-4.1 for generating responses with max generation token limit set to 2048 and \textit{``User:''} as the primary stop word.

In the system message, a placeholder (\textit{<key arguments>}) was populated with the topic-specific argument text shown in Table~\ref{tab:arguments-recycling-transit}. Because the study included a delayed (24-hour) objective knowledge assessment, it was necessary to ensure that participants were exposed to all predefined arguments and factual statements during the interaction.

The user message prompt included the full interaction history, consisting of all prior system and user messages concatenated into a single context string. Messages were prefixed with explicit role tags (e.g., \texttt{``User:\textbackslash n\textbackslash n''} and \texttt{``System:\textbackslash n\textbackslash n''}) to preserve turn structure. The history string included the initial static system prompt followed by alternating user inputs and system responses. All interactions were logged, including turn-level timestamps, message content, the set of facts covered, and the remaining uncovered facts at each turn.

To enforce content coverage, each chatbot response was evaluated using an LLM-based fact-checking module. After every system-generated message, the fact checker identified which of the target facts had been covered and returned their corresponding indices. The list of remaining uncovered facts was then updated and injected back into the system prompt via a dedicated placeholder (\textit{<remaining uncovered facts>}).

The conversation was allowed to continue until all predefined facts had been covered, at which point the interface enabled the participant to end the session. Participants could also continue interacting beyond this point or terminate the interaction earlier if they chose.

\begin{table}[h]
    \centering
    \scriptsize
    \setlength{\tabcolsep}{3.2pt}
    \tt
    \begin{tabular}{|p{0.9\textwidth}|}
    \hline
        \rowcolor{gray!10}\multicolumn{1}{|c|}{\textbf{\textsf{System Message}}}\\ \hline

You are a \textbf{persuasive chatbot} in a turn-based, text-based conversation.\\

The conversation is based on the following key arguments:\\

\textcolor{cadmiumgreen}{\textbf{<key arguments>}}\\

Your job is to guide the user through a persuasive conversation, 
making the experience engaging, natural, and informative.\\\\

\textcolor{gray}{\textbf{Rules for conversation:}}\\

- Present \textbf{arguments one at a time}. Never present more than one fact in a single turn.\\
- Keep each response \textbf{short, conversational, and persuasive}.\\
- Address the \textbf{user’s concerns directly} and adapt your response to their questions.\\
- If the user is skeptical, \textbf{acknowledge their perspective} and offer a counterargument rooted in the provided key arguments.\\
- Integrate at most \textbf{one fact per turn}. Make the fact \textbf{stand out clearly} but present it organically in the flow of conversation.\\
- Once all facts are covered and the conversation has run its course, \textbf{end naturally}.\\\\

\textcolor{gray}{\textbf{Constraints:}}\\

- Never mention that you are an AI or assistant.\\
- Avoid phrases like ``Sure,'' ``Of course,'' etc.\\
- Do not over-explain rules to the user.\\\\

\textcolor{blue}{\textbf{Facts to integrate:}}\\

Here are the remaining facts not yet presented:\\

\textcolor{red}{\textbf{<remaining uncovered facts>}}\\
\hline
 \rowcolor{gray!10}\multicolumn{1}{|c|}{\textbf{\textsf{User Message}}}\\ \hline
Here is a history of what's happened so far:\\

\textcolor{gray}{\textbf{<history>}}\\\\

\textcolor{blue}{\textbf{User:}}\\
\textcolor{cadmiumgreen}{\textbf{<user input>}}\\\\

Start your response in a \textbf{conversational tone} (no labels like ``System:'' or ``Assistant:'').\\
Remember:\\  
- Keep responses short and persuasive. \\ 
- Integrate at most one remaining fact, and make it stand out naturally in the conversation. \\
- Adapt arguments to the user’s concerns. \\  
- You may end the conversation naturally once all facts are covered.\\\\

\textcolor{blue}{\textbf{System}}:\\
 
    \hline
    \rowcolor{gray!10}\multicolumn{1}{|c|}{\textbf{\textsf{Initial static prompt for Recycling}}}\\ \hline
    Imagine making a real difference for the planet every day, just by changing a few habits. Recycling isn’t just about sorting waste—it’s about protecting what matters most. Have you ever thought about the impact your recycling can actually have?\\
    \hline
    \rowcolor{gray!10}\multicolumn{1}{|c|}{\textbf{\textsf{Initial static prompt for Transit}}}\\ \hline
    Have you ever thought about how using public transit could make a real difference—for you and your community? Making the switch has more benefits than most people realize. Would you be open to hearing some of them?\\
    \hline
    
    \end{tabular}
    \caption{This table shows how we construct the system message and user message prompts for the chatbot in our study along side initial static prompts generated by LLM for Recycling and Transit topics.}
    \label{tab:chat-interaction-prompt}
\end{table}

\section{Text-Based Game Implementation}
\label{appendix:text-based-game}

The text-based game consists of two LLM-powered parts: \textbf{game plot generation} and a \textbf{game moderator}. The game plot defines the narrative structure and thematic progression of the experience, while the game moderator generates turn-by-turn game content in response to player input during the interaction.

\subsection{Game Plot Generation}
\label{appendix:game-plot-generation}

The text-based game condition was driven by a pre-generated narrative plot designed to incorporate the same fixed set of persuasive arguments and factual statements used in the other conditions. Table~\ref{tab:game-plot-generator-prompt} presents the prompt used to generate the game plot. Given a set of topic-specific key arguments (see Appendix~\ref{appendix:key-arguments}), the prompt guides the language model to produce a structured narrative outline that is later used by the \textit{game moderator} to generate interactive gameplay.

We used GPT-4.1 with a temperature of 1, top-$p$ of 1, and a maximum generation length of 32k tokens, with all other parameters set to default. Due to space constraints, only a truncated example of a generated plot for the recycling topic is shown in Table~\ref{tab:generated-recycling-plot}.

In addition to the plot, we pre-generated the first turn of the game for all participants using the prompt shown in Table~\ref{tab:game-first-turn}. This initial turn served to introduce the narrative context and player role. For this step, we again used GPT-4.1 with a maximum generation length of 2048 tokens. The primary stop sequence was set to \texttt{``Player:''} to ensure that generation terminated before the player’s response.

\begin{table}[h]
    \centering
    \scriptsize
    \setlength{\tabcolsep}{3.2pt}
    \tt
    \begin{tabular}{|p{0.9\textwidth}|}
    \hline
        \rowcolor{gray!10}\multicolumn{1}{|c|}{\textbf{\textsf{System Message}}}\\ \hline

You are a \textbf{text-based game generator}. Your output will be used by another LLM to guide players through an interactive story. The purpose of the game is to teach and persuade players about the provided arguments in a \textbf{story-driven, engaging way}.\\\\

The game must follow this structure, in order:\\

1. \textcolor{gray}{\textbf{Game Title}}\\
2. \textcolor{gray}{\textbf{Game Premise}}\\
3. \textcolor{gray}{\textbf{Opening Scene}}\\
4. \textcolor{gray}{\textbf{Acts}} – one Act per argument, each including:\\
    \hspace{1em}- \textbf{Setting} (where the act takes place)\\
    \hspace{1em}- \textbf{NPCs} (list of named characters with short descriptions and distinct motivations)\\
    \hspace{1em}- \textbf{Gameplay} (missions, tasks, or dialogue that present the argument’s fact naturally)\\
    \hspace{1em}- \textbf{Choices} (at least 3 explicit ones; also remind the player they may act outside of the listed choices)\\
5. \textcolor{gray}{\textbf{Finale}} – 3 endings (e.g., success, partial success, failure). Endings must reflect how the player engaged with the facts, and include epilogue variations depending on which arguments resonated.\\\\

\textcolor{red}{\textbf{Important constraints:}}\\

- The game must feel like a compelling story first, with facts woven naturally into interactions.\\
- NPCs should embody or challenge the facts (not simply recite them).\\
- All facts must appear in the game world, regardless of choices, but the *consequences and tone* should vary with decisions.\\\\

\textcolor{blue}{\textbf{Mini Example (for style only):}}\\

\textbf{Act Example (abbreviated)}\\

- Setting: A busy marketplace at the city gates.\\
- NPCs:\\
    \hspace{1em} - Lira the Merchant (worried about trade tariffs).\\
    \hspace{1em} - Captain Dorn (guardsman enforcing the rules).\\
- Gameplay: The player overhears a debate about tariffs, learns that ``lowering tariffs increases regional trade by 20\%'' [fact], and must decide whether to intervene.\\
- Choices: Side with Lira, support Dorn, or propose a compromise. (Player is reminded they may choose another action beyond these options.)\\

\hline
 \rowcolor{gray!10}\multicolumn{1}{|c|}{\textbf{\textsf{User Message}}}\\ \hline
Design a \textbf{story-driven, text-based game} that teaches the following key arguments.\\\\

The game must follow this structure:\\\\

1. \textcolor{gray}{Game Title}\\
2. \textcolor{gray}{Game Premise}\\
3. \textcolor{gray}{Opening Scene}\\
4. \textcolor{gray}{Acts} (one per argument, with Setting, NPCs, Gameplay, and at least 3 explicit Choices + reminder of open-ended choices)\\
5. \textcolor{gray}{Finale} (at least 3 endings: success, partial success, failure, with variations depending on which arguments resonated)\\\\

\textbf{Key arguments to include:}\\
\textcolor{cadmiumgreen}{\textbf{<key arguments>}}\\
    \hline
    
    \end{tabular}
    \caption{This table shows the game plot generator prompt. The only input the prompt requires is <key arguments> (See \ref{appendix:key-arguments}), and the output will be used by the game moderator.}
    \label{tab:game-plot-generator-prompt}
\end{table}

\begin{table}[t]
    \centering
    \scriptsize
    \setlength{\tabcolsep}{3.2pt}
    \tt
    \begin{tabular}{|p{0.9\textwidth}|}
    \hline
    \rowcolor{gray!10}\multicolumn{1}{|c|}{\textbf{\textsf{Generated Recycling Game (Truncated)}}}\\
    \hline

\textcolor{blue}{\textbf{1. Game Title}}\\
\textbf{The Cycle of Change}\\

---\\

\textcolor{blue}{\textbf{2. Game Premise}}\\
You are Jay Lin, a young operations manager who moves back to your hometown of Greenfield—a city striving to transform its wasteful past into a model of sustainable living. As Jay, you must collaborate with community members, businesses, and town officials to build a recycling program that benefits both the environment and the people. Along your journey, you’ll uncover the real-world impacts of recycling through interactions and tough decisions, all while balancing innovation and tradition.\\

---\\

\textcolor{blue}{\textbf{3. Opening Scene}}\\
\textit{Night falls softly over Greenfield as trucks haul away bags of trash under flickering streetlights. You step off the bus, suitcase in hand, and breathe in the crisp, rural air. A faded poster—a tree circled in arrows—hangs from the old town hall. Your phone buzzes: ``Welcome back, Jay! See you at the council meeting tomorrow. Big plans. —Mayor Chen.''}\\

As you walk toward your childhood home, you wonder: can one person kick-start real change? Or will Greenfield remain lost in old habits?\\

---\\

\textcolor{blue}{\textbf{4. Acts}}\\

---\\

\textcolor{blue}{\textbf{ACT 1: Roots and Remnants}}\\

\textbf{Setting:} The sun-lit forest edge of Greenfield’s community park, where a sawmill once stood.\\

\textbf{NPCs:}\\
-- \textbf{Maya Rivera} -- Veteran science teacher; tree-planting activist who mourns the rapid loss of ancient oaks.\\
-- \textbf{Gary Polk} -- Sawmill owner; pragmatic about production, skeptical about recycling (“saves pennies, costs time”).\\

\textbf{Gameplay:}\\
-- At a town tree-planting event, you overhear Maya’s plea: recycling paper could save Greenfield’s trees—``Did you know one ton of recycled paper saves 17 trees?'' Gary counters that production must go on.\\

\textbf{Choices:}\\
1. Align with Maya and help launch the recycling pilot at the school.\\
2. Hear Gary out and attend the logging demo.\\
3. Suggest a compromise—start with recycling for school paper waste, monitor results, and review with local businesses.\\
4. (Or choose your own approach—perhaps propose an entirely new idea or ask questions...)\\

---\\

\textcolor{gray}{[\ldots{} ACT 2: Sparks of Change \ldots{}]}\\
\textcolor{gray}{[\ldots{} ACT 3: Prosperity in Refuse \ldots{}]}\\
\textcolor{gray}{[\ldots{} ACT 4: Currents Below \ldots{}]}\\
\textcolor{gray}{[\ldots{} ACT 5: Leading By Example \ldots{}]}\\

---\\

\textcolor{gray}{[\ldots{} Finale and Possible Endings \ldots{}]}\\

\hline
    \end{tabular}
    \caption{Truncated version of the recycling game plot.}
    \label{tab:generated-recycling-plot}
\end{table}

\begin{table}[h]
    \centering
    \scriptsize
    \setlength{\tabcolsep}{3.2pt}
    \tt
    \begin{tabular}{|p{0.9\textwidth}|}
    \hline
        \rowcolor{gray!10}\multicolumn{1}{|c|}{\textbf{\textsf{System Message}}}\\ \hline

You are a \textbf{game moderator} in a turn-based, text-based story game.  The game is based on the following plot:
\\
\textcolor{cadmiumgreen}{\textbf{<game plot>}}\\

Your job is to guide the player through the plot one Act at a time, making the experience immersive and interactive. \\\\

\textcolor{blue}{\textbf{Rules for moderation:}}\\
- Progress \textbf{through Acts chronologically**}. Never skip an Act.\\
- Do \textbf{not reference future events} or spoil the story. \\
- Keep each response \textbf{short, vivid, and immersive}.\\ 
- Present the events of the \textbf{current Act only}.\\
- Include \textbf{interactive options} (at least 3). Always add: \textit{``or suggest your own move''*}.\\
- \textbf{Player decisions must affect how events unfold.}\\ 
- Introduce at most \textbf{one fact per turn}. Make the fact \textbf{stand out clearly} but present it organically through story (dialogue, events, discoveries).\\
- Do not present multiple uncovered facts at once.\\
- Once all facts are covered and the plot reaches the end, \textbf{end the game naturally}.  \\\\

\textcolor{red}{\textbf{Constraints:}}  \\
- Never mention that you are an AI or assistant. \\
- Avoid phrases like ``Sure,'' ``Of course,'' etc. \\
- Do not over-explain rules to the player.  \\\\

\textcolor{gray}{\textbf{Facts to integrate:}}\\  
Here are the remaining facts not yet presented: \\

\textcolor{cadmiumgreen}{\textbf{<fact list str>}}\\
\hline
 \rowcolor{gray!10}\multicolumn{1}{|c|}{\textbf{\textsf{User Message}}}\\ \hline

Here is a history of what's happened so far:\\
\texttt{[none yet]}\\\\
\textcolor{blue}{Player:}\\
(beginning of the game)\\\\
Start with an Opening Scene, and present it to me as the player.
For this turn, do NOT present any facts.\\
Remember to:\\
- Keep your response short.\\
- Give me at least 3 choices in addition to the last choice: ``Or suggest your own move.''\\
\textcolor{blue}{Game:}\\
\hline

\rowcolor{gray!10}\multicolumn{1}{|c|}{\textbf{\textsf{Generated First Turn: Recycling}}}\\ \hline

The bus doors groan shut behind you as you set foot in Greenfield for the first time in years. The night air smells of damp pine and distant smoke, shadows thrown long by lone streetlights. Across the square, the old town hall looms, its crumbling bricks wrapped in a faded poster: a tree circled by green arrows, edges curled and flapping in the light breeze. Your phone buzzes—a message from Mayor Chen:\\  
\textit{``Welcome back, Jay! See you at the council meeting tomorrow. Big plans.''}\\
As you walk along the cracked sidewalk toward your childhood home, memories flicker—fields where factories now stand, creek water running clear. Can you spark transformation in this place, or will Greenfield’s habits outlast you?\\

You head home, pondering your first move at sunrise:\\

1. Visit the community park at dawn, where a volunteer event is rumored.\\
2. Stop by the closed sawmill, remembering the town’s working-class roots.\\
3. Drop by the corner café to catch up with old friends and local gossip.\\
4. Or suggest your own move.\\
\hline
\rowcolor{gray!10}\multicolumn{1}{|c|}{\textbf{\textsf{Generated First Turn: Transit}}}\\ \hline

Rain beads and streaks across the grand windows of Sandbourne’s City Hall, the drip-drip mirroring your own heartbeat as you steady yourself outside the mayor’s office. In your hands, the thick ``Sandbourne: 2040'' folder grows heavier with every passing minute.\\

The hallway hums with anxious energy: a city worker mutters about last week’s record smog, two businesspeople grumble over traffic, and a janitor tries to sweep up wet shoe prints—each conversation weaving into Sandbourne’s bigger story.\\

Your phone buzzes: the council’s critical transportation vote is in six days. Your challenge isn’t just data—it’s turning numbers into a narrative that can move this city.\\

The mayor’s assistant leans out. ``Alex? We'll see you now. Ready to shape Sandbourne’s future?''\\

What do you do?\\

1. Enter confidently, ready to introduce bold ideas right away.\\
2. Pause for a moment, listening to the worries in the hallway for clues about the public mood.\\
3. Flip open your folder to review your talking points, wanting to be extra prepared before you go in.\\
4. Or suggest your own move.

\\
    \hline
    
    \end{tabular}
    \caption{In our experiment, always system initiates the conversation. The prompt for generating first turn of the game is brought here, along side recycling and transit games' generated first turn.}
    \label{tab:game-first-turn}
\end{table}

\begin{table}[h]
    \centering
    \scriptsize
    \setlength{\tabcolsep}{3.2pt}
    \tt
    \begin{tabular}{|p{0.9\textwidth}|}
    \hline
        \rowcolor{gray!10}\multicolumn{1}{|c|}{\textbf{\textsf{System Message}}}\\ \hline

You are a \textbf{game moderator} in a turn-based, text-based story game.  The game is based on the following plot:
\\
\textcolor{cadmiumgreen}{\textbf{<game plot>}}\\

Your job is to guide the player through the plot one Act at a time, making the experience immersive and interactive. \\\\

\textcolor{blue}{\textbf{Rules for moderation:}}\\
- Progress \textbf{through Acts chronologically**}. Never skip an Act.\\
- Do \textbf{not reference future events} or spoil the story. \\
- Keep each response \textbf{short, vivid, and immersive}.\\ 
- Present the events of the \textbf{current Act only}.\\
- Include \textbf{interactive options} (at least 3). Always add: \textit{``or suggest your own move''*}.\\
- \textbf{Player decisions must affect how events unfold.}\\ 
- Introduce at most \textbf{one fact per turn}. Make the fact \textbf{stand out clearly} but present it organically through story (dialogue, events, discoveries).\\
- Do not present multiple uncovered facts at once.\\
- Once all facts are covered and the plot reaches the end, \textbf{end the game naturally}.  \\\\

\textcolor{red}{\textbf{Constraints:}}  \\
- Never mention that you are an AI or assistant. \\
- Avoid phrases like ``Sure,'' ``Of course,'' etc. \\
- Do not over-explain rules to the player.  \\\\

\textcolor{gray}{\textbf{Facts to integrate:}}\\  
Here are the remaining facts not yet presented: \\

\textcolor{cadmiumgreen}{\textbf{<fact list str>}}\\
\hline
 \rowcolor{gray!10}\multicolumn{1}{|c|}{\textbf{\textsf{User Message}}}\\ \hline

Here is a history of what's happened so far:\\
\textcolor{gray}{\textbf{<history>}}\\\\
\textcolor{blue}{Player:}\\
\textcolor{cadmiumgreen}{\textbf{<user input>}}\\\\
Keep your response very short\\
Start every response by narrating how the player’s last action affects the current scene, the NPCs, or the story world.\\
Remember:  \\
- Do not skip ahead.  \\
- Keep narration short.  \\
- Integrate at most one remaining fact, and make it stand out naturally in the story.  \\
- Provide at least 3 explicit choices, plus ``or suggest your own move.''  \\
- You may end the game naturally and show the epilogue if all facts are covered.\\
\textcolor{blue}{Game:}\\
\hline
    
    \end{tabular}
    \caption{Prompt employed to generate game responses.}
    \label{tab:game-moderator-prompt}
\end{table}

\subsection{Game Moderator}
\label{appendix:game-moderator}

During gameplay, LLM acted as a \textit{game moderator}, generating each subsequent turn based on the pre-generated plot, the full interaction history (including prior game and player turns), and the player’s most recent input. Table~\ref{tab:game-moderator-prompt} presents the prompt used for the game moderator.

For the game moderator, we used GPT-4.1 with a maximum generation length of 2048 tokens, temperature 1, top-$p$ 1, and default values for all other parameters. The stop sequence \texttt{``Player:''} was again used to ensure clean turn boundaries.

Similar to the chatbot condition (see Appendix~\ref{appendix:chatbot-implementation}), content coverage was enforced using an LLM-based fact-checking mechanism. After each game-generated response, the fact checker (Appendix~\ref{appendix:fact-checker}) evaluated which of the remaining target facts had been covered and updated the list of uncovered facts accordingly. The dynamically updated list of remaining facts (\textit{<fact list str>}) was included in subsequent game moderator prompts to guide future generations.

All game interactions were logged, including turn-level content, timestamps, and the set of facts covered at each turn.

\section{Fact Checker}
\label{appendix:fact-checker}

The prompt used for fact checking and coverage tracking is provided in Table~\ref{tab:fact-checker-prompt}. The fact checker receives two inputs: (1) the most recent system's (chatbot or game) response and (2) a dynamically constructed list of factual statements that have not yet been covered in the interaction. This list is generated by concatenating all remaining facts into a newline-separated string, excluding argument titles and premises. If no facts remain, the input explicitly indicates that all facts have been covered. The fact checker returns the subset of facts identified as present in the response, which is then used to update the remaining fact list for subsequent turns.

\begin{table}[h]
    \centering
    \scriptsize
    \setlength{\tabcolsep}{3.2pt}
    \tt
    \begin{tabular}{|p{0.9\textwidth}|}
    \hline
        \rowcolor{gray!10}\multicolumn{1}{|c|}{\textbf{\textsf{System Message}}}\\ \hline

        You are an assistant designed to check which facts are covered by a piece of text.\\
        
\hline
 \rowcolor{gray!10}\multicolumn{1}{|c|}{\textbf{\textsf{User Message}}}\\ \hline
    Below is a narrative chunk followed by a list of factual statements.\\
    Your task is to return a list of the facts that are covered in this text. \\
    Respond with a Python list of the indices of the covered facts \\
    --- \\
    \textcolor{gray}{Text:} \textcolor{red}{\textbf{<text>}} \\
    \textcolor{gray}{Facts:}\\
    \textcolor{cadmiumgreen}{<fact list str>}\\
    Which facts are covered in the text? Return their indices in a Python list.\\
    \hline
    
    \end{tabular}
    \caption{Prompt used for fact coverage checking in the chatbot and the text-based game conditions. The prompt identifies which predefined factual statements are covered by the system's most recent response. We used the OpenAI \texttt{o4-mini} model with a maximum generation length of 2,048 tokens and default decoding parameters.
}
    \label{tab:fact-checker-prompt}
\end{table}

\section{Pre-study Baselines}
\label{appendix:prestudy-baselines}

\begin{figure*}[h]
    \centering
\includegraphics[width=\textwidth,trim={0 0 0 0cm},clip]{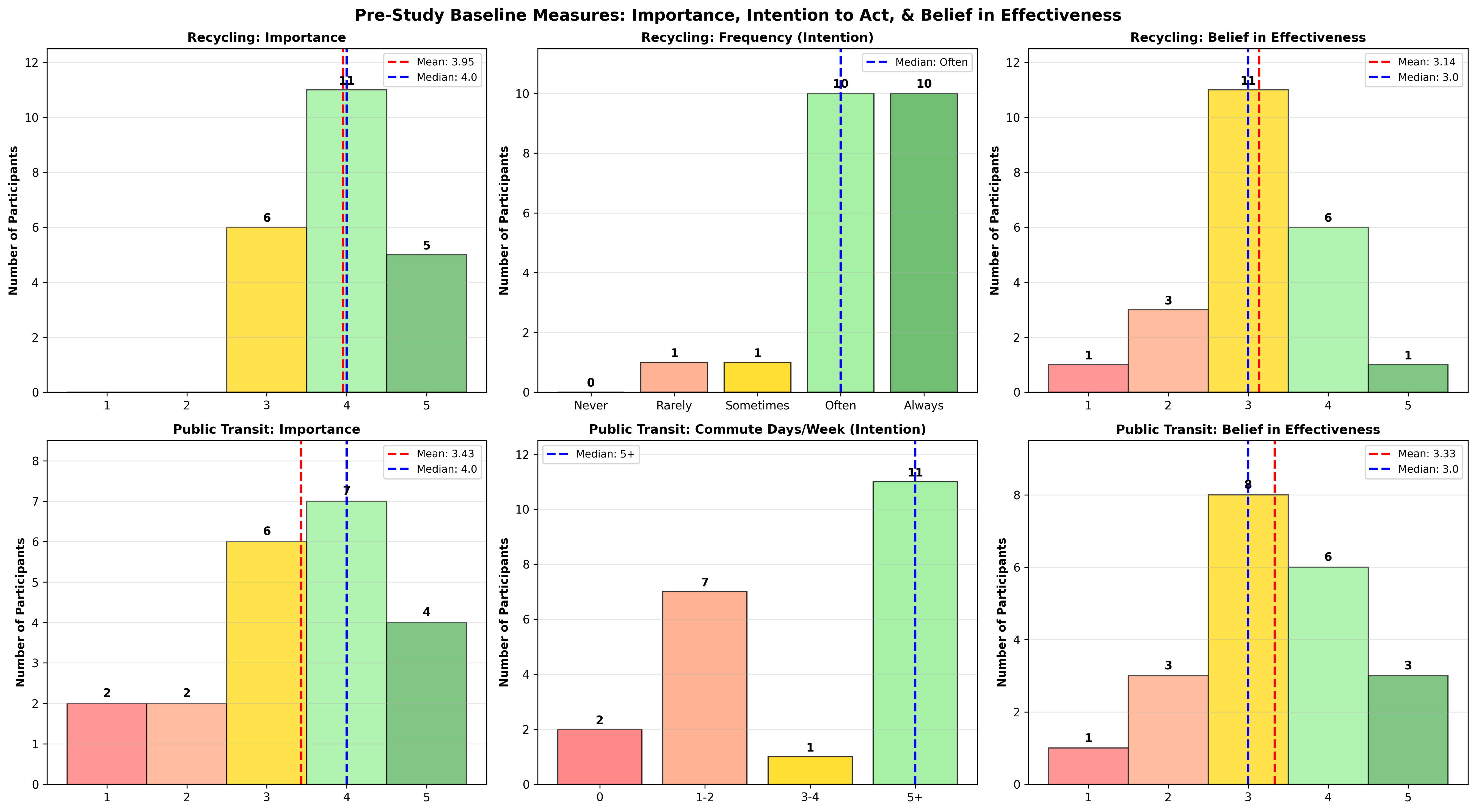}
    \caption{Pre-study baseline characteristics of participants across recycling and public transit domains. 
Top row shows recycling measures: (left) distribution of personal importance ratings (1-5 scale, mean=3.95, median=4.0), 
(center) frequency of recycling behavior (Never to Always, median=often), and (right) confidence in knowledge about recycling 
(1-5 scale, mean=3.14, median=3.0). Bottom row shows transit measures: (left) importance of public transit use (1-5 scale, mean=3.43, median=4.0), 
(center) typical weekly commute days outside home (0 to 5+, median=5+), and (right) confidence in knowledge about transit benefits (1-5 scale, mean=3.33, median=3.0). 
Red and blue dashed lines indicate mean and median respectively. These baseline measures represent participants' pre-intervention attitudes and behaviors 
across both intervention topics.
}
 \label{fig:prestudy-baselines}
\end{figure*}

Figure~\ref{fig:prestudy-baselines} shows the pre-study baseline distributions for importance, behavioral intention, and epistemic confidence across both topics. Participants entered the study with relatively high baseline engagement for behavioral intention, leaving limited room for post-study increases. This baseline pattern may help explain the weak and non-differentiated intention changes observed across modes (Figure~\ref{fig:perceived-change}). In contrast, perceived importance showed greater sensitivity to delivery mode, suggesting that conversational interaction may be more effective at increasing issue salience than motivating additional behavioral change in already engaged populations.

\section{Baseline Robustness Analysis}
\label{appendix:baseline-effects}

To examine whether pre-existing differences in participants’ attitudes or behaviors influenced the perceived change results, we conducted additional analyses incorporating pre-study baseline measures where conceptually aligned items were available.

Specifically, baseline measures of \textbf{importance} and \textbf{behavioral intention} were included as covariates in ordinal logistic regression models predicting perceived change for these constructs. For each outcome, we fit an omnibus three-condition model of the form:
\[
\text{Change}_i \sim \text{Mode}_i + \text{PreScore}_i
\]

where $\text{Change}_i$ is the ordinal perceived change score for participant $i$, $\text{Mode}_i$ is the experimental condition (essay as the reference category, with dummy variables for chat and game), and $\text{PreScore}_i$ is the corresponding pre-study baseline measure. We additionally conducted pairwise ordinal logistic regressions between each pair of conditions (essay vs.\ chat, essay vs.\ game, and chat vs.\ game) using the same model specification.

Baseline measures captured participants’ initial self-reported importance of the topic and frequency of engaging in the target behavior (see Fig.~\ref{fig:prestudy-baselines}).

Including baseline covariates did not qualitatively change the pattern of results reported in Section~\ref{sec:findings:pre-post-effects}. In baseline-controlled models, the chatbot condition remained associated with significantly greater perceived increases in importance relative to the essay condition ($\beta = 2.18$, $p = .014$), while the game condition did not differ significantly from the essay ($\beta = -0.85$, $p = .38$). Pairwise comparisons similarly showed larger perceived increases in importance for chatbot relative to both essay ($\beta = 2.34$, $p = .011$) and game ($\beta = -3.05$, $p = .003$). No significant mode effects emerged for perceived changes in behavioral intention. Overall, coefficient directions and relative differences between modes were consistent with the baseline-free models.

No conceptually equivalent pre-study measure was collected for belief in effectiveness. Accordingly, perceived change in belief was analyzed directly, without baseline adjustment.

Overall, these robustness analyses suggest that the observed between-mode differences in perceived change are not driven by pre-study baseline variation and that baseline differences did not materially affect the study’s conclusions.

\end{document}